\documentclass[a4paper, 11pt]{article}
\usepackage[dvipdfmx]{graphicx}
\usepackage{amsmath,amsthm,amssymb,mathrsfs,mathtools,color,yhmath}
\usepackage[titletoc,title]{appendix}
\usepackage{enumitem,kantlipsum}
\usepackage{subcaption}
\usepackage{multirow}

\usepackage[numbers ,sort&compress]{natbib}
\newcommand{\rnc}{\renewcommand}

\rnc{\Im}{{\textrm{Im}\,}}
\rnc{\Re}{{\textrm{Re}\,}}

\newcommand{\pa}[1]{{\mathrm{p}_{#1}}}
\newcommand{\paprime}[1]{{\mathrm{p}^\prime_{#1}}}

\newcommand{\co}{\mathrm{c}}
\newcommand{\po}{\mathrm{xy}}
\newcommand{\x}{\mathrm{x}}
\newcommand{\y}{\mathrm{y}}
\newcommand{\Lo}{\mathrm{L}}
\newcommand{\E}{\mathrm{E}}

\newcommand{\gs}{\propto}

\newcommand{\cx}{{\mathcal{CX}}}
\newcommand{\cz}{{\mathcal{CZ}}}
\newcommand{\cs}{{\mathrm{CS}}}
\newcommand{\cpi}{{\mathrm{C}\pi}}
\newcommand{\cpihalf}{{\mathrm{C}\frac{\pi}{2}}}

\newcommand{\ctrl}{{\mathrm{ctrl}}}
\newcommand{\tgt}{{\mathrm{tgt}}}

\newcommand{\cnot}{{\mathrm{CNOT}}} 

\newcommand{\nodagger}{{\vphantom{\dagger}}}

\DeclareMathOperator{\tr}{tr}

\theoremstyle{definition}

\numberwithin{equation}{section}

\begin{document}

\title{
    Toward fault-tolerant quantum computation exploiting quantum spatial distribution and gauge symmetry
}

\author{
    Ryo Asaka\thanks{E-mail: asaka@rs.tus.ac.jp}
    \\\\
    \textit{Department of Physics,
        Tokyo University of Science,}\\
    \textit{Kagurazaka 1-3, Shinjuku-ku, Tokyo 162-8601, Japan} \\
    \\\\
    \\
}

\date{\today}

\maketitle

\begin{abstract}
    We explore how the integrated use of
    quantum spatial distribution (QSD), or more specifically, a superposition of both spin and position states of particles,
    and gauge symmetry (GS) revealed by Poulin's stabilizer formalism enhances quantum error correction.
    The study employs $3+2$ particles on nested squares proposed in the companion paper (arXiv:2504.07941),
    where three of them encode Shor's nine-qubit code and the remaining two detect errors in this code through their spin state measurements.
    The first result is that the GS offers resilience against three types of noise acting on a particle:
    arbitrary decoherence of its spin or position state, and dephasing of both states, which completely or partly destroys its QSD.
    To show that, we formulate a noise model unifying the above noise sources and prove the correctability of this unified model under our error-correcting scheme.
    The second result is that the QSD, together with the GS, provides architectural flexibility, allowing us to stack the error-correcting systems both vertically and horizontally.
    Indeed, we present implementations of the error detection (stabilizer measurement),
    logical Hadamard and Toffoli gates, and a quantum adder with the required interactions only between nearest-neighbor and next-nearest-neighbor particles.
    Here, our treatment of the dynamics of particles, each having spin and position degrees of freedom,
    under nontrivial noise and gate operations indicates that the stabilizer formalism
    is a powerful tool for describing quantum many-body dynamics.
\end{abstract}

\maketitle
\section{Introduction}\label{Section: Introduction}
In quantum mechanics, superposition is a unique property whereby a single particle can simultaneously occupy multiple positions,
reflecting its wave-like nature. 
Here, we refer to such a phenomenon as quantum spatial distribution (QSD).
The most well-known evidence of the QSD is perhaps the interference pattern projected onto the screen beyond a double-slit wall~\cite{feynman1965feynman,zeilinger1988single,carnal1991young,tonomura1989demonstration}.
Indeed, such a pattern is a consequence of a single particle passing through both slits simultaneously and interfering with itself as if it were two particles~\cite{feynman1965feynman,tonomura1989demonstration}.

A recent experiment has successfully manipulated the QSD of a single particle (electron, referred to as a flying qubit) over two separate regions,
suggesting that this phenomenon could be directly exploited in quantum computing~\cite{yamamoto2012electrical, bautze2014theoretical,bauerle2018coherent,duprez2019macroscopic,takada2019sound,edlbauer2022semiconductor}.
In particular, the experiment~\cite{yamamoto2012electrical} has demonstrated an arbitrary unitary transformation
on the superposition of two states $|0\rangle_{\mathrm{y}}$ and $|1\rangle_{\mathrm{y}}\in\mathbb{C}^2$ that represent the particle's two possible positions.
The $X$-axis Bloch rotations are implemented via tunnel coupling between the two regions, whereas $Z$-axis rotations are realized through the Aharonov-Bohm effect.
Two flying qubits become entangled through the two-particle scattering (Coulomb interaction) \cite{wang2023coulomb,fletcher2023time,ubbelohde2023two,ouacel2025electronic}.

One advantage of the QSD is the ability to send a single particle to multiple points simultaneously,
whereby the particle can undergo multiple conditional gate operations based on each destination~\cite{cerf1998optical}.
For example, quantum random access memory, a prominent theoretical proposal,
exploits this phenomenon to simultaneously query $O(2^n)$ memory cells with only $O(n^2)$ time complexity~\cite{giovannetti2008quantum,giovannetti2008architectures,hann2019hardware,chen2021scalable,wang2025quantum}.
A particle (photon) simultaneously reaches multiple cells through the QSD, and undergoes gate operations
to encode a qubit state $|0\rangle$ or $|1\rangle\in\mathbb{C}^2$ into its spin (polarization) state~\cite{giovannetti2008quantum}, with each cell attached to a leaf of a full binary tree of depth $n$,
and a quantum switch placed at each node to direct the particle to both branches in a quantum superposition.

The QSD---which, from here on, also encompasses the superposition of a particle's spin states---can also offer advantages in reducing the number of particles in quantum computation and eliminating time-dependence in gate operations.
As an example of the former, one can represent a multi-qubit state using fewer particles than qubits, 
by encoding quantum information into both spin and position states of the particles~\cite{singh2021universal,chawla2023multi,asaka2025quantum}.
Another example is the elimination of the aforementioned $2^{n}-1$ quantum switches (qutrits or three-level particles)
of the quantum random access memory, by having each memory-querying particle select its destination at each bifurcation based on its spin state~~\cite{asaka2021quantum,asaka2023two2,de2025resource}.
As an example of the latter advantage, researchers have proposed autonomous scattering-based quantum computation in which a particle (a quantum walker) encodes qubit information 
into its position state~~\cite{childs2009universal,childs2013universal,bao2015universal} 
or both spin and position states~\cite{asaka2023two1,asaka2023two2} and undergoes computational operations by simply passing through the corresponding gates.

A key nontrivial theoretical challenge for ensuring the practicality and feasibility of the QSD-based computational architectures,
which have been proposed or are yet to be proposed, is incorporating a quantum error-correcting scheme into them.
For this, we would need to develop a novel scheme that protects against noise affecting the position and spin states of a particle, i.e., the multi-qubit state within a particle.
Dephasing noise, which we refer to as partial loss or complete vanishing of the QSD,
would also constitute the dominant source of errors in such architectures.
Moreover, the error-correcting scheme would also need to have architectural flexibility so that the advantages of the QSD are preserved.

Toward addressing this nontrivial challenge, 
the present study takes a first step; we employ a simple quantum error-correcting model to demonstrate the fundamental advantages offered by 
the integrated use of the QSD and the gauge symmetry (GS) revealed by the stabilizer formalism~\cite{gottesman1996class,gottesman1997stabilizer} extended by Poulin~\cite{poulin2005stabilizer}.
Specifically, we demonstrate that the GS provides resilience against the dominant noise inherent in QSD-based architectures.
Furthermore, we show that the integrated use of the QSD and the GS provides the architectural flexibility that preserves the advantages of the QSD.
Here, the notion of the GS is based on the unified and generalized error-correcting theorem known as {\it operator quantum error correction}~\cite{kribs2005unified};
a brief historical background on this topic is deferred to Sec.~\ref{Section: Error Correction}.

\begin{figure}[t]
    \centering
    \begin{minipage}[b]{0.29\columnwidth}
        \centering
        \includegraphics[width=3.cm]{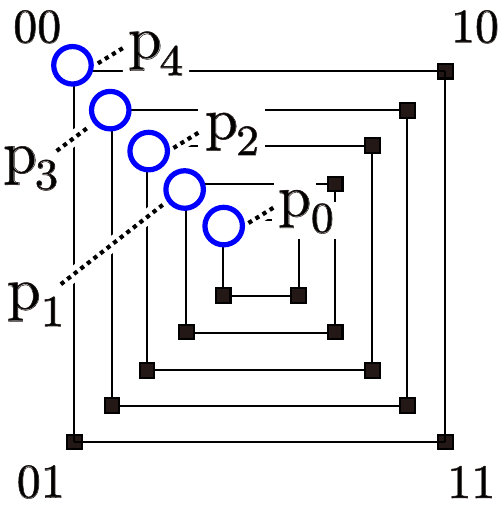}
        \subcaption{}
    \end{minipage}
    \begin{minipage}[b]{0.69\columnwidth}
        \centering
        \includegraphics[width=9.cm]{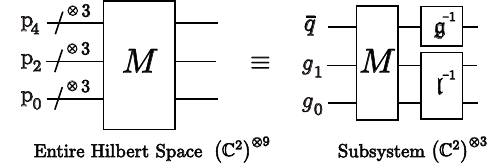}
        \subcaption{}
    \end{minipage}
    \caption{
    (a) Three physical particles $\{\pa{0},\pa{2},\pa{4}\}$ and two ancillary particles $\{\pa{1},\pa{3}\}$.
    We specify four vertices of each square as $00$, $10$, $11$, and $01$.
    (b) Schematic of the gauge symmetry [Eqs.~\eqref{eq: gauge symmetry} and \eqref{eq: gauge symmetry (operator)}],
    where $M$ is any operation in $\mathrm{End}((\mathbb{C}^2)^{\otimes 9})$ and $\mathfrak{g}$ (resp.~$\mathfrak{l}$) is any operation in the gauge algebra generated by the gauge (resp.~logical) transformations and stabilizer generators in Table~\ref{table: stabilizer generators}.
    The effect that the virtual logical qubit $\bar{q}$ (resp.~the virtual gauge qubits $\{g_0, g_1\}$) receives from any operation $M$ acting on the three physical particles is $M$ modulo the corresponding algebra.
    }
    \label{figure: gauge symmetry}
\end{figure}

The demonstration of the fundamental advantages of the QSD and the GS is performed using a simple QSD-based architecture proposed in the companion paper~\cite{asaka2025quantum}, 
$3+2$ particles on five nested squares [Fig.~\ref{figure: gauge symmetry}~(a)].
This model encodes Shor's nine-qubit error-correcting code~\cite{shor1995scheme} into the three particles, each of which has spin and position states in $\mathbb{C}^2$ and $(\mathbb{C}^2)^{\otimes 2}$, respectively.
The remaining two particles are used to detect errors in these three particles.
Here, Shor's code is known to possess the GS, as revealed by Poulin~\cite{poulin2005stabilizer}.

The first result of the demonstration is that the GS within the present error-correcting model offers resilience against three types of dominant noise in our QSD-based architecture:
(i) unified spin-related noise, (ii) unified position-related noise, and (iii) partial loss or complete vanishing of the QSD
(the noise (i) also depends on the particle's position state).
Specifically, we formulate a noise model that unifies the above three noise types,
and show that our error-correcting scheme eliminates the influence of this unified noise acting on one of the three particles.
From this formulation, we can explicitly verify that the absence of the GS means the unified noise falls outside the correctable class.

The second result is that the QSD combined with the GS provides a form of architectural flexibility, namely stacking flexibility, for scalable quantum computing within quantum error correction.
Specifically, we demonstrate that the logical Hadamard and Toffoli gates, which together can simulate universal quantum computation~\cite{shi2002both,aharonov2003simple}, are implemented with two nested-squares-systems communicating
through nearest-neighbor (and next-nearest-neighbor) interactions between outermost-to-outermost (and outermost-to-next-outermost) particles.
In addition, we show that a quantum adder is implemented by the stacking of the systems horizontally and vertically,
which reflects the architectural flexibility offered by the integrated use of the QSD and the GS.

This paper is organized as follows.
In Sec.~2, we formulate our model of $3 + 2$ particles on the nested squares and provide an overview of the following sections.
In Sec.~3, we discuss our error-correcting mechanism together with the GS.
In Sec.~4, we demonstrate the resilience against the aforementioned three types of noise, which is one of the two main results of this paper.
In Sec.~5, we show that the QSD, together with the GS, offers the stacking flexibility for the quantum computation, which is the second main result.
We also briefly estimate the failure probability of the error correction in the Appendix.

\section{Model Setup}\label{Section: Model Setup}
To demonstrate that the GS offers resilience against the dominant noise inherent in a QSD-based architecture and
the QSD, together with the GS, offers the architectural flexibility in quantum error correction,
we employ $3$+$2$ particles allocated to each of the five nested squares (indexed from the inside as $i=0, 1, \cdots , 4$).
In this section, we first formulate the states (QSDs) of the particles, and then introduce gate operations for the particles,
along with a brief overview of Secs.~\ref{Section: Error Correction}-\ref{Section: Scalability Enhancement}.

Note that, as opposed to the quantum-walk-based error-correcting model proposed in the companion paper~\cite{asaka2025quantum}, 
the particles evolve in discrete time under gates that act independently on each state, outside the quantum walk framework.
This setting solidifies the effectiveness of the integrated use of the QSD and the GS,
independently of the ultrafast realization by taking the continuous limit of the quantum walk operations.

\subsection{Physical and Ancillary Particles}\label{Subsection: }
We refer to the three and two particles on the nested squares as the ``physical'' and ``ancillary'' particles, respectively.
Here, the physical particles $\{\pa{0}, \pa{2}, \pa{4}\}$, each residing in the 1st, 3rd, or 5th square from the innermost,
form a QSD that encodes Shor's nine-qubit code~\cite{shor1995scheme}.
Meanwhile, the remaining two ancillary particles $\{\pa{1}, \pa{3}\}$ are used for the stabilizer measurement explained in Sec.~\ref{Subsection: Stabilizer measurement}.
Note that, in Sec.~\ref{Section: Noise Resilience Enhancement}, 
we will show the resilience against dominant noise acting on a physical particle offered by the GS inherent in Shor's code, revealed by using Poulin's stabilizer formalism~\cite{poulin2005stabilizer}.

Each particle has spin (coin) and position states, represented by $|c\rangle_\co \in \mathbb{C}^2\ (c\in\{0,1\})$ and $|xy\rangle_\po\in(\mathbb{C}^2)^{\otimes 2}\ (x,y\in\{0,1\})$, respectively.
Here, $|c\rangle_\co$ denotes the up- or down-spin state, while $|xy\rangle_\po$ specifies the vertex where the particle is located.
Sometimes, we decompose the position state as $|xy\rangle_\po = |x\rangle_\x |y\rangle_\y$.
In addition, to identify the particle, we enclose these states in parentheses and append the subscript $\pa{i}\ (0\leq i\leq 4)$ on the right.
In particular, based on the above rules, we define the QSDs that encode information $0$ or $1$ as follows:
\begin{align}
    |0\rangle_\Lo := \frac{|+\rangle_\Lo + |-\rangle_\Lo}{\sqrt{2}},\     
    |1\rangle_\Lo := \frac{|+\rangle_\Lo - |-\rangle_\Lo}{\sqrt{2}},
    \label{eq: logical states}
\end{align}
where
\begin{align}
    |\pm\rangle_\Lo := \bigotimes_{i\in\{0,2,4\}}\Bigl(\frac{|0\rangle_\co|00\rangle_\po \pm |1\rangle_\co|11\rangle_\po}{\sqrt{2}}\Bigr)_{\pa{i}}.
    \label{eq: logical states (X-basis)}
\end{align}
We will discuss the theoretical background of such a definition in Sec.~\ref{Subsection: Gauge Symmetry}.

\subsection{Gate operations}\label{Subsection: Gate operations}
We first summarize gate operations required in our error-correcting scheme
to demonstrate that the GS offers the resilience to the aforementioned three types of noise inherent in QSD-based architectures.
Second, we introduce gate operations to implement quantum logical gates
to demonstrate that the integrated use of the QSD and the GS offers the stacking flexibility.
These gate operations constitute our entire computational model, and no additional gate operation will appear in the subsequent sections.

Before introducing the gate operations, we note that the following three notational rules apply to operators throughout the present paper:
(i)  The subscripts attached to an operator specify a particle and its state that this operator acts on.
Namely, $(U_\co)_\pa{i}$, $(U_\x)_\pa{i}$, or $(U_\y)_\pa{i}$ denotes that some operator $U\in \mathrm{End}(\mathbb{C}^2)$
acts on the spin, $x$-axis, or $y$-axis position state of  the particle $\pa{i}$, respectively.
(ii) We denote the position-state dependence of a spin-state operation by a superscript as follows:
\begin{align}
    U_\co^{[\{xy\}]} := \sum_{xy} U_\co \otimes |xy\rangle\langle xy|_\po
    \label{eq: position-dependent spin operation}
\end{align}
for $x,y\in\{0,1\}$,
e.g., $X_\co^{[00,11]} = X_\co \otimes |00\rangle\langle 00|_\po + X_\co \otimes |11\rangle\langle 11|_\po$, where $X$ is the Pauli $X$ gate.
(iii) We implicitly assume that the identity acts on all states whose actions are not explicitly represented,
e.g., under $X_\co^{[00,11]}$, a particle whose position state is $|01\rangle_\po$ or $|10\rangle_\po$ implicitly undergoes the identity $I_\co$
(though we may explicitly represent it for clarity).

First, the error-correcting scheme, i.e., stabilizer measurement in Sec.~\ref{Section: Error Correction}, consists of three types of gate operations:
the Pauli $X$ and $Z$ gates acting on the spin state ($X_\co$ and $Z_\co$); the Hadamard gate acting on the spin, $x$- or $y$-axis position state ($H_\co$, $H_\x$, and $H_\y$);
and the controlled-NOT gate acting on spin states of the neighboring physical and ancillary particles ($\cnot_\co$).
Here, under this controlled gate, the Pauli $X$ gate acts on the spin state of $\pa{j}$
if and only if the spin state of $\pa{i}$ is a $-1$ eigenstate of $Z_\co$, and both particles share the same position state:
\begin{align}
    (\cnot_\co)_{\pa{i},\pa{j}} :=  \sum_{xy}\sum_{p=0}^1 \left(\frac{I_\co + (-1)^p Z_\co}{2}\otimes|xy\rangle\langle xy|_\po\right)_\pa{i} (X^p_\co \otimes|xy\rangle\langle xy|_\po)_\pa{j},
    \label{eq: definition of cnot c}
\end{align}
where $j=i \pm 1$ holds for the nearest-neighbor controlled-NOT gate.

The stabilizer measurement employs the above gates to map the eigenvalue of physical particle $\pa{i}$ under the Pauli operator ($Z_\co Z_\x I_\y$, $Z_\co I_\x Z_\y$, or $X_\co X_\x X_\y$) for $i\in\{0,2,4\}$
to the spin state of the neighboring ancillary particle $\pa{j}$ for $j\in\{1,3\}$, as illustrated in Fig.~\ref{figure: fig/stabilizer_measurement}~(b).
These Pauli operators are among the stabilizer generators in Table~\ref{table: stabilizer generators}.
Here, in Eqs.~\eqref{eq: stabilizer measurement (impl 1)}--\eqref{eq: stabilizer measurement (impl 3)},  we will explicitly verify the validity of the mapping procedure.

Second, the implementation of quantum computation requires $\cnot_\co$ [Eq.~\eqref{eq: definition of cnot c}] acting on two next-nearest-neighboring particles in addition to the above gate operations.
All of the gates are employed for operations across two nested-squares-systems,
i.e., for implementing two of the three primitive logical gates that realize the Hadamard and Toffoli gates: $\cx$ [Eq.~\eqref{eq: CX definition}] and $\cz$ [Eq.~\eqref{eq: CZ definition}].
Specifically, given another nested-square system $\{\paprime{i}|0\leq i\leq 4\}$ with the two systems aligned horizontally,
we require nearest-neighbor $(\cnot_\co)_{\pa{4},\paprime{4}}$ and next-nearest-neighbor $(\cnot_\co)_{\pa{3},\paprime{4}}$ as
\begin{align}
    \begin{array}{c}
        \includegraphics[height=2.5cm]{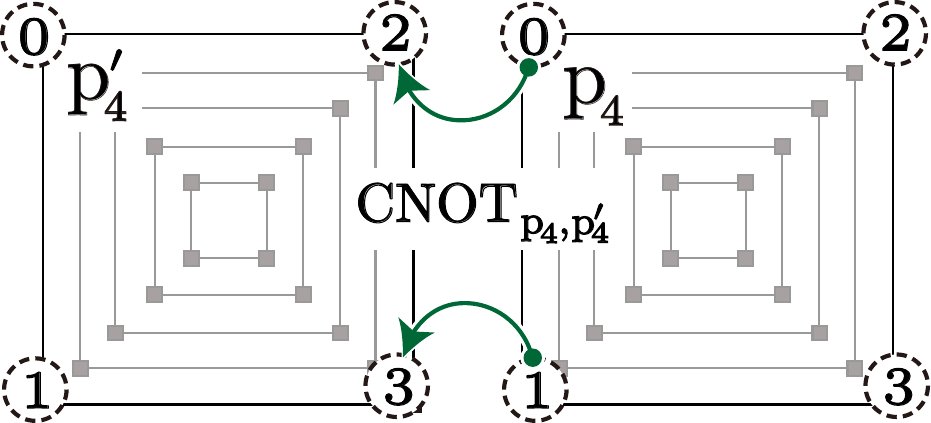},\quad
        \includegraphics[height=2.5cm]{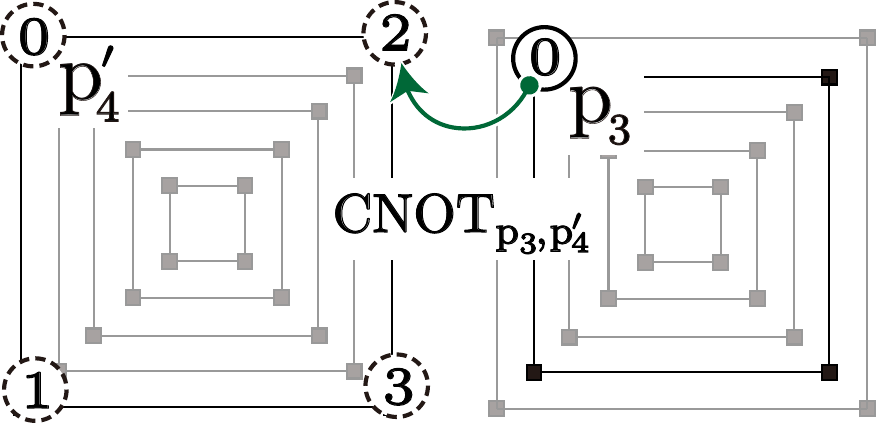},
    \end{array}
    \label{eq: }
\end{align}
along with next-nearest-neighbor $(\cnot_\co)_{\pa{1},\pa{3}}$ between the ancillary particles.
Here, each circle with a dashed (resp.~solid) line denotes the state of a particle (resp.~an actual particle) whose spin-state is arbitrary and whose position state is specified by the number inside the circle.
Conversely, if the two systems are aligned vertically,
the two primitive logical gates $\cx$ and $\cz$ can be implemented using nearest-neighbor $(\cnot_\co)_{\pa{i},\paprime{i}}$ for $i\in\{0,2,4\}$ without any next-nearest-neighbor interaction.

In addition, to implement the remaining primitive logical gate $\sqrt{\cx}$ [Eq.~\eqref{eq: CXroot definition}],
we require the following three types of nearest-neighbor interactions between the two outermost particles of the two systems (i.e., $i,j=4$) instead of $(\cnot_\co)_{\pa{4},\paprime{4}}$:
{\allowdisplaybreaks
\begin{align}
    (\cs_\co)_{\pa{i},\pa{j}}      & :=  \sum_{xy}\sum_{p=0}^1 \left(\frac{I_\co + (-1)^p Z_\co}{2}\otimes|xy\rangle\langle xy|_\po\right)_\pa{i} (S_\co \otimes|xy\rangle\langle xy|_\po)^p_\pa{j},
    \label{eq: definition of nearest-neighbor cs}                                                                                                                                                    \\
    (\cpi_\co)_{\pa{i},\pa{j}}     & :=  \sum_{xy}\sum_{p=0}^1 \left(Z_\co\otimes|xy\rangle\langle xy|_\po\right)_\pa{i} (I_\co \otimes|xy\rangle\langle xy|_\po)^p_\pa{j},
    \label{eq: definition of nearest-neighbor cpi}                                                                                                                                                   \\
    (\cpihalf_\co)_{\pa{i},\pa{j}} & :=  \sum_{xy}\sum_{p=0}^1 \left(S_\co\otimes|xy\rangle\langle xy|_\po\right)_\pa{i} (I_\co \otimes|xy\rangle\langle xy|_\po)^p_\pa{j},
    \label{eq: definition of nearest-neighbor cshalf}
\end{align}
}
where $S:= |0\rangle\langle 0| + i |1\rangle\langle 1|$ ($\cpi$ and $\cpihalf$ can be written as Eq.~\eqref{eq: other form of cpi and cpihalf}).

\newcommand{\colsep}{\hspace{4pt}}
\begin{table}[tp]
    \caption{
        (Left) Stabilizer generators and (Right) Logical and gauge transformations.
    }
    \label{table: stabilizer generators}
    \begin{tabular}{cc}
        \begin{minipage}[t]{0.48\columnwidth}
            \centering
            \begin{tabular}{c|c@{\colsep}c@{\colsep}c@{\colsep}|c@{\colsep}c@{\colsep}c@{\colsep}|c@{\colsep}c@{\colsep}c@{\colsep}}
                      & \multicolumn{3}{c}{$\pa{4}$} & \multicolumn{3}{c}{$\pa{2}$} & \multicolumn{3}{c}{$\pa{0}$}                                                         \\\hline
                \\[-1em]
                $s_0$ & $I_\co$                      & $I_\x$                       & $I_\y$                       & $Z_\co$ & $Z_\x$ & $I_\y$ & $Z_\co$ & $Z_\x$ & $I_\y$ \\
                $s_1$ & $Z_\co$                      & $Z_\x$                       & $I_\y$                       & $Z_\co$ & $Z_\x$ & $I_\y$ & $I_\co$ & $I_\x$ & $I_\y$ \\
                $s_2$ & $I_\co$                      & $I_\x$                       & $I_\y$                       & $Z_\co$ & $I_\x$ & $Z_\y$ & $Z_\co$ & $I_\x$ & $Z_\y$ \\
                $s_3$ & $Z_\co$                      & $I_\x$                       & $Z_\y$                       & $Z_\co$ & $I_\x$ & $Z_\y$ & $I_\co$ & $I_\x$ & $I_\y$ \\
                $s_4$ & $I_\co$                      & $I_\x$                       & $I_\y$                       & $X_\co$ & $X_\x$ & $X_\y$ & $X_\co$ & $X_\x$ & $X_\y$ \\
                $s_5$ & $X_\co$                      & $X_\x$                       & $X_\y$                       & $X_\co$ & $X_\x$ & $X_\y$ & $I_\co$ & $I_\x$ & $I_\y$ \\
            \end{tabular}
        \end{minipage}
        \begin{minipage}[t]{0.48\columnwidth}
            \centering
            \begin{tabular}{c|c@{\colsep}c@{\colsep}c@{\colsep}|c@{\colsep}c@{\colsep}c@{\colsep}|c@{\colsep}c@{\colsep}c@{\colsep}}
                                & \multicolumn{3}{c}{$\pa{4}$} & \multicolumn{3}{c}{$\pa{2}$} & \multicolumn{3}{c}{$\pa{0}$}                                                         \\\hline
                \\[-1em]
                $\bar{Z}$       & $Z_\co$                      & $Z_\x$                       & $Z_\y$                       & $Z_\co$ & $Z_\x$ & $Z_\y$ & $Z_\co$ & $Z_\x$ & $Z_\y$ \\
                $\bar{X}$       & $X_\co$                      & $X_\x$                       & $X_\y$                       & $X_\co$ & $X_\x$ & $X_\y$ & $X_\co$ & $X_\x$ & $X_\y$ \\
                $\bar{Z}_{g_0}$ & $Z_\co$                      & $Z_\x$                       & $I_\y$                       & $Z_\co$ & $Z_\x$ & $I_\y$ & $Z_\co$ & $Z_\x$ & $I_\y$ \\
                $\bar{X}_{g_0}$ & $X_\co$                      & $I_\x$                       & $X_\y$                       & $X_\co$ & $I_\x$ & $X_\y$ & $X_\co$ & $I_\x$ & $X_\y$ \\
                $\bar{Z}_{g_1}$ & $Z_\co$                      & $I_\x$                       & $Z_\y$                       & $Z_\co$ & $I_\x$ & $Z_\y$ & $Z_\co$ & $I_\x$ & $Z_\y$ \\
                $\bar{X}_{g_1}$ & $X_\co$                      & $X_\x$                       & $I_\y$                       & $X_\co$ & $X_\x$ & $I_\y$ & $X_\co$ & $X_\x$ & $I_\y$ \\
            \end{tabular}
        \end{minipage}
    \end{tabular}

    \vspace{1.5em}  

    \caption{
        Elements of the recovery operation in Eq.~\eqref{eq: recovery operation}.
    }
    \label{table: recovery operations}
    \centering
    \begin{tabular}{@{}c@{\hspace{2em}}c@{}}
        \begin{tabular}{cc|c}
            $m_5$ & $m_4$ & $\sigma_Z$                   \\\hline
            $0$   & $0$   & \rule{0pt}{5mm}$\bar{I}$     \\
            $0$   & $1$   & $(Z_\co\ I_\x\ I_\y)_\pa{0}$ \\
            $1$   & $0$   & $(Z_\co\ I_\x\ I_\y)_\pa{4}$ \\
            $1$   & $1$   & $(Z_\co\ I_\x\ I_\y)_\pa{2}$ \\
        \end{tabular}
         &
        \begin{tabular}{cccc|c}
            $m_3$ & $m_2$ & $m_1$ & $m_0$ & $\sigma_X$                   \\\hline
            $0$   & $0$   & $0$   & $0$   & \rule{0pt}{5mm}$\bar{I}$     \\
            $0$   & $1$   & $0$   & $1$   & $(X_\co\ I_\x\ I_\y)_\pa{0}$ \\
            $0$   & $0$   & $0$   & $1$   & $(I_\co\ X_\x\ I_\y)_\pa{0}$ \\
            $0$   & $1$   & $0$   & $0$   & $(I_\co\ I_\x\ X_\y)_\pa{0}$ \\
        \end{tabular}
        \\[3.5em]
        \begin{tabular}{cccc|c}
            $m_3$ & $m_2$ & $m_1$ & $m_0$ & $\sigma_X$                                  \\\hline
            $1$   & $1$   & $1$   & $1$   & \rule{0pt}{5mm}$(X_\co\ I_\x\ I_\y)_\pa{2}$ \\
            $0$   & $0$   & $1$   & $1$   & $(I_\co\ X_\x\ I_\y)_\pa{2}$                \\
            $1$   & $1$   & $0$   & $0$   & $(I_\co\ I_\x\ X_\y)_\pa{2}$                \\
        \end{tabular}
         &
        \begin{tabular}{cccc|c}
            $m_3$ & $m_2$ & $m_1$ & $m_0$ & $\sigma_X$                                  \\\hline
            $1$   & $0$   & $1$   & $0$   & \rule{0pt}{5mm}$(X_\co\ I_\x\ I_\y)_\pa{4}$ \\
            $0$   & $0$   & $1$   & $0$   & $(I_\co\ X_\x\ I_\y)_\pa{4}$                \\
            $1$   & $0$   & $0$   & $0$   & $(I_\co\ I_\x\ X_\y)_\pa{4}$                \\
        \end{tabular}
    \end{tabular}
\end{table}

\newpage
\section{Error Correction}\label{Section: Error Correction}
In this section, we first provide theoretical background on the logical QSDs $|0\rangle_\Lo$ and $|1\rangle_\Lo$ [Eq.~\eqref{eq: logical states}] and the GS
revealed by the framework of Poulin's stabilizer formalism~\cite{poulin2005stabilizer}.
Second, we discuss the mechanism and implementation of the stabilizer measurement by which we prepare the logical QSD and correct errors in these states.
The description of this mechanism provides intuition for the resilience to the three types of dominant noise, although an explicit demonstration of it will be presented in the next section.

Historically, a subsystem structure, which underlies the GS [cf.~Fig.~\ref{figure: gauge symmetry}~(b)],
was first recognized by Ref.~\cite{knill2000theory} within error-correcting codes as a key notion for encoding and protecting quantum information.
Ref.~\cite{zanardi2000stabilizing} subsequently extended this viewpoint by revealing the same structure underlying the frameworks of decoherence-free subspace encoding~\cite{duan1997preserving,duan1998reducing,zanardi1997error,lidar1998decoherence}
and noise suppression~\cite{vitali1999using,viola1998dynamical}.
The operator quantum error correction~\cite{kribs2005unified} was then introduced to unify these frameworks, and later reformulated within the stabilizer formalism~\cite{gottesman1996class,gottesman1997stabilizer,dauphinais2024stabilizer} by Poulin.
The research toward the fault-tolerant quantum computation exploiting the subsystem structure~\cite{kempe2001theory} is also crucial.
Notably, a recent study~\cite{egan2021fault} has experimentally demonstrated fault-tolerant logical operations of the Bacon-Shor code~\cite{bacon2006operator,aliferis2007subsystem} that is based on the operator quantum error correction,
as in the present error-correcting scheme.

\subsection{Gauge Symmetry}\label{Subsection: Gauge Symmetry}
Our error-correcting mechanism arises from the introduction of the six stabilizer generators $s_0$--$s_5$ listed in Table~\ref{table: stabilizer generators}~(Left).
As discussed below, the logical QSDs $|0\rangle_\Lo$ and $|1\rangle_\Lo$ in Eq.~\eqref{eq: logical states} and the GS inherent in this mechanism are described by
the logical and gauge transformations ($\{\bar{Z}, \bar{X}\}$ and $\{\bar{Z}_{g_i}, \bar{X}_{g_i}|i\in\{0,1\}\}$) in Table~\ref{table: stabilizer generators}~(Right).

We first define the simultaneous eigenspace of these generators $s_0$--$s_5$ as the {\it gauge subsystem}, into which single-qubit information is encoded.
Since each of $s_0$--$s_5$ divides the entire Hilbert space of the three physical particles ($\mathcal{H}:=(\mathbb{C}^{2})^{\otimes 9}$) into its $\pm 1$ eigenspaces,
the gauge subsystem is isomorphic to the three-qubit Hilbert space $\left(\mathbb{C}^2\right)^{\otimes 3}$ ($=\mathbb{C}^{2^{9-6}}$).
Here, we regard our gauge subsystem as consisting of three virtual qubits: a ``logical'' qubit $\bar{q}$ and two ``gauge'' qubits $g_0$ and $g_1$ that carry the GS.

The logical qubit hosts the encoded quantum information that is protected from noise by our error-correcting mechanism.
We define $Z$- and $X$-basis states of this virtual qubit as $\pm 1$ eigenstates of the logical Pauli operators $\bar{Z}$ and $\bar{X}$, respectively, which are listed in Table~\ref{table: stabilizer generators}~(Right).
Indeed, the logical QSDs in Eq.~\eqref{eq: logical states} are $+1$ (resp.~$-1$) eigenstates of $\bar{Z}$,
storing the noise-protected information $|0\rangle\in\mathbb{C}^2$ (resp.~$|1\rangle$) in the logical qubit.
Here, they are also simultaneous $+1$ eigenstates of $s_0$--$s_5$.
In addition, $|\pm \rangle_\Lo$ in Eq.~\eqref{eq: logical states (X-basis)} is the $\pm 1$ eigenstate of $\bar{X}$, which is also stabilized by $s_0$--$s_5$.

Meanwhile, the gauge qubits, whose $Z$- and $X$-basis states are specified by the gauge transformations $\{\bar{Z}_{g_i}, \bar{X}_{g_i}|i\in\{0,1\}\}$ in Table~\ref{table: stabilizer generators}~(Right),
do not carry logical information but rather induce the GS~\cite{poulin2005stabilizer}.
As illustrated by the schematic in Fig.~\ref{figure: gauge symmetry}~(b),
the information stored in the logical qubit experiences the same effect under any operations that are equivalent modulo the gauge algebra $\mathcal{G}$ generated by the gauge transformations and the stabilizer generators.
This symmetry follows from the commutation of the logical Pauli operators $\{\bar{Z}, \bar{X}\}$ with both $\{\bar{Z}_{g_i}, \bar{X}_{g_i}|i\in\{0,1\}\}$ and $s_0$--$s_5$,
meaning $\mathcal{G}$ does not affect $\bar{Z}$ and $\bar{X}$ eigenstates.

We denote the GS by the symbol $\gs$, which means that the effect of the LHS operator on the logical qubit equals that of the RHS modulo $\mathcal{G}$:
\begin{align}
    M\left(\alpha |0\rangle_\Lo + \beta|1\rangle_\Lo\right) \gs M \mathfrak{g}^{-1} \left(\alpha|0\rangle_\Lo+\beta |1\rangle_\Lo\right),\
    \alpha,\beta\in\mathbb{C},\ \forall \mathfrak{g}\in\mathcal{G},
    \label{eq: gauge symmetry}
\end{align}
where $M\in \mathrm{End}((\mathbb{C}^2)^{\otimes 9})$ is any operator acting on the three physical particles.
We also express the above symmetry at the operator level as
\begin{align}
    M  \gs M\mathfrak{g}^{-1}.
    \label{eq: gauge symmetry (operator)}
\end{align}

As will be discussed in Secs.~\ref{Section: Noise Resilience Enhancement} and \ref{Section: Scalability Enhancement},
the symmetry expands the scope of correctable errors, yielding the resilience against the three types of dominant noise in our architecture.
Furthermore, as will be shown in Sec.~\ref{Section: Scalability Enhancement}, the symmetry allows us to localize the logical operations onto outermost particles,
yielding the stacking flexibility.

\subsection{Stabilizer Measurement}\label{Subsection: Stabilizer measurement}
In this section, we discuss the mechanism of stabilizer measurement, i.e., the measurement of the $s_0$--$s_5$ eigenvalues,
for preparing the logical QSDs $|0\rangle_\Lo$ or $|1\rangle_\Lo$ [Eq.~\eqref{eq: logical states}], and correcting errors in these states.
Note that we represent the measured $s_i$ eigenvalue by $m_i\in\{0,1\}$ $(0\leq i\leq 5)$ as $(-1)^{m_i}$,
and the overall outcome of the stabilizer measurement as
\begin{align}
    m = m_5 m_4 m_3 m_2 m_1 m_0 \in \{0,1\}^6.
    \label{eq:  overall outcome of the stabilizer measurement}
\end{align}

\paragraph{Mechanism}
Through the stabilizer measurement together with the measurement of the $\bar{Z}$ eigenvalue,
we can prepare either the logical state $|0\rangle_\Lo$ or $|1\rangle_\Lo$, typically accompanied by flipping errors that change some stabilizer eigenvalues from $+1$ to $-1$.
Although our QEC requires no immediate correction of these errors as we mention in the final paragraph of this section and explicitly demonstrate in Sec.~\ref{Subsection: Unified Noise Model},
we can remove them by applying the following recovery operation:
\begin{align}
    \sigma^\dagger_m := \sigma^\dagger_Z (m_5, m_4) \sigma^\dagger_X (m_3, m_2, m_1, m_0),
    \label{eq: recovery operation}
\end{align}
where $\sigma_Z (m_5, m_4)$ and $\sigma_X (m_3, m_2, m_1, m_0)$ are defined in Table~\ref{table: recovery operations}.
In addition, although the measurement of the $\bar{Z}$ eigenvalue yields either $+1$ or $-1$ randomly,
meaning that $|0\rangle_\Lo$ or $|1\rangle_\Lo$ emerges,
we can change them arbitrarily by applying $X_\co$ to all physical particles,
exploiting the following GS:
\begin{align}
    \bigotimes_{i\in\{0,2,4\}} (X_\co)_\pa{i} = \left(\bar{X}_{g_0}\bar{X}_{g_1}\right)\cdot\bar{X} \gs \bar{X},
    \label{eq: }
\end{align}
where we use $\bar{X}_{g_0}\bar{X}_{g_1}\in\mathcal{G}$.

To correct an error arising in the logical QSDs, we also perform the stabilizer measurement and record the measured outcomes.
As will be detailed in Sec.~\ref{Section: Noise Resilience Enhancement},
the error, i.e., an undesired operation in the logical qubit [cf.~Fig.~\ref{figure: gauge symmetry}~(b)], can be corrected
if the noise source falls within the unified noise model [Eq.~\eqref{Subsection: Unified Noise Model}] containing all three types of dominant noise mentioned in the introduction.
The reason why the stabilizer measurement functions against such noise is that, intuitively, each operator element $E^{(n)}$ in Eq.~\eqref{eq: unified quantum noise}
that acts as a nontrivial error on the logical qubit
collapses to the trivial flipping errors in Table~\ref{table: recovery operations} as
\begin{align}
    E^{(n)} = \sum_{m\in\{0,1\}^6} \mathfrak{g}_{n,m} \sigma_m \rightarrow \mathfrak{g}_{n,m} \sigma_m \gs \sigma_m,
    \label{eq: }
\end{align}
where $\mathfrak{g}_{n,m}\in\mathcal{G}$ for $n\in\mathbb{N}_0$.
Here, the measurement outcome $m\in\{0,1\}^6$ informs us that these flipping errors $\sigma_m$ remain in the logical qubit.

In practice, rather than applying a recovery operation $\sigma^\dagger_m$ immediately after each stabilizer measurement,
we perform error correction collectively after $N$ rounds of stabilizer measurements.
Namely, we can correct accumulated errors in the system by applying the operations $\prod^{N-1}_{i=0} \sigma^\dagger_{m^i}$
determined by the history of the measurement outcomes $\{m^i\in \{0,1\}^6\mid 0\leq i\leq N-1\}$.
This recovery strategy is known as a {\it Pauli frame update}~\cite{knill2005quantum} (or {\it offline error correction}~\cite{egan2021fault}), and its validity in our scheme is explicitly justified by Eq.~\eqref{eq: pauli frame update} in the next section.

\paragraph{Implementation}
Our implementation of the stabilizer measurement consists of three stages: eigenvalue measurements of $(s_0,s_1)$, $(s_2,s_3)$, and $(s_4,s_5)$.
In each stage of the stabilizer measurement,
we map the $s_k$ and $s_{k+1}$ ($k\in\{0,2,4\}$) eigenvalues to the spin states of the ancillary particles $\pa{1}$ and $\pa{3}$, respectively,
and measure these spin states, yielding outcomes $m_k$ and $m_{k+1}\in\{0,1\}$.

\begin{figure}
    \vspace{0.3cm}
    \centering
    \includegraphics[width=14cm]{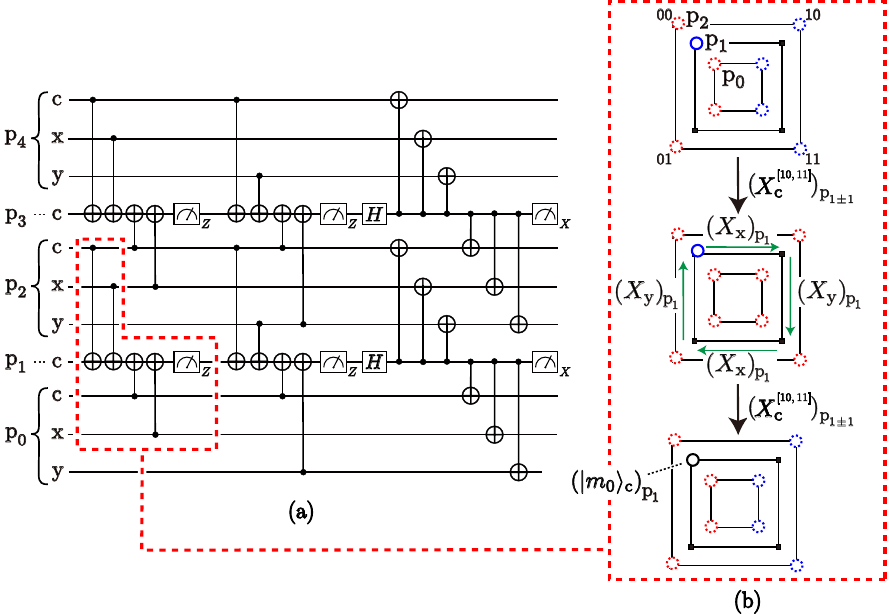}
    \caption{
    (a) Scheme of the Stabilizer measurement, i.e., measurement of $s_0$-$s_5$ eigenvalues.
    Before measuring the spin state of ancillary particle $\pa{1}$ (resp.~$\pa{3}$) at stage $k \in\{0,2,4\}$, the $s_k$ (resp.~$s_{k+1}$) eigenvalue is mapped onto this spin state.
    Here, each mapping consists of submappings,
    each of which transfers the $Z_\co Z_\x I_\y$, $Z_\co I_\x Z_\y$ or $X_\co X_\x X_\y$ eigenvalue of physical particle $\pa{j\pm 1}$
    onto the spin state of ancillary particle $\pa{j}$.
    (b) Procedure for $s_0$ eigenvalue mapping ($(|0\rangle_\co)_{\pa{1}} \rightarrow (|m_0\rangle_\co)_{\pa{1}}$),
    where $m_0\in\{0,1\}$ denotes that the eigenvalue of $s_0$ is $(-1)^{m_0}$.
    The color of circles represents the particles' spin state, blue for $|0\rangle_\co$ and red for $|1\rangle_\co$.
    [i]~Each dashed circle denotes a $-1$ eigenstate of physical particle $\pa{0}$ (resp.~$\pa{2}$) under the operator $Z_\co Z_\x I_\y$.
    [ii]~Those spin states are all turned into $|1\rangle_\co$ by $X_\co^{[10,11]}$, and thereby activate the nearest-neighbor $\cnot_{\pa{0},\pa{1}}$ (resp.~$\cnot_{\pa{2},\pa{1}}$)
    applied before each tunneling.
    [iii]~As a result, the spin state of $\pa{1}$ becomes $|1\rangle_\co$ if and only if exactly one of $\pa{0}$ and $\pa{2}$ is in a $-1$ eigenstate under $Z_\co Z_\x I_\y$,
    meaning the system is in a $-1$ eigenstate under the stabilizer generator $s_0$, and $|0\rangle_\co$ otherwise .
    }
    \label{figure: fig/stabilizer_measurement}
\end{figure}

Specifically, each mapping consists of submappings of the eigenvalues of the physical particles $\pa{j\pm 1}$
under a Pauli operator ($Z_\co Z_\x I_\y$, $Z_\co I_\x Z_\y$, or $X_\co X_\x X_\y$) onto the spin state of the ancillary particle $\pa{j}$ for $j\in\{1,3\}$,
as the circuit Fig.~\ref{figure: fig/stabilizer_measurement}~(a).
Here, procedures for such submappings are derived as
\begin{align}
     & \sum^1_{p=0} \left(\frac{I_\co I_\x I_\y + (-1)^p Z_\co Z_\x I_\y}{2}\right)_\pa{j\pm 1} (X_\co)^p_{\pa{j}}
    \notag                                                                                                                                                                                                       \\
     & = \left(X_\co^{[10,11]}\right)_{\pa{j\pm 1}} \left((X_\y)_{\pa{j}} (\cnot_\co)_{\pa{j\pm 1},\pa{j}} (X_\x)_{\pa{j}} (\cnot_\co)_{\pa{j\pm 1},\pa{j}}\right)^2 \left(X_\co^{[10,11]}\right)_{\pa{j\pm 1}},
    \label{eq: stabilizer measurement (impl 1)}
\end{align}
\begin{align}
     & \sum^1_{p=0} \left(\frac{I_\co I_\x I_\y + (-1)^p Z_\co I_\x Z_\y}{2}\right)_\pa{j\pm 1} (X_\co)^p_{\pa{j}}
    \notag                                                                                                                                                                                                       \\
     & = \left(X_\co^{[11,01]}\right)_{\pa{j\pm 1}} \left((X_\y)_{\pa{j}} (\cnot_\co)_{\pa{j\pm 1},\pa{j}} (X_\x)_{\pa{j}} (\cnot_\co)_{\pa{j\pm 1},\pa{j}}\right)^2 \left(X_\co^{[11,01]}\right)_{\pa{j\pm 1}},
    \label{eq: stabilizer measurement (impl 2)}
\end{align}
\begin{align}
     & \sum^1_{p=0} \left(\frac{I_\co I_\x I_\y + (-1)^p X_\co X_\x X_\y}{2}\right)_\pa{j\pm 1} (X_\co)^p_{\pa{j}}
    \notag                                                                                                                                                                                                        \\
     & = \left(\left(H_\x H_\y\cdot Z_\co^{[01,10]} \right)_\pa{j\pm 1}\left(H_\co\right)_\pa{j}\right)
    \notag                                                                                                                                                                                                        \\
     & \qquad  \left((X_\y)_{\pa{j}} (\cnot_\co)_{\pa{j},\pa{j\pm 1}} (X_\x)_{\pa{j}} (\cnot_\co)_{\pa{j},\pa{j\pm 1}}\right)^2 \left(\left(Z_\co^{[01,10]}\cdot H_\x H_\y\right)_\pa{j\pm 1}\left(H_\co\right)_\pa{j}\right).
    \label{eq: stabilizer measurement (impl 3)}
\end{align}
Namely, as illustrated in Fig.~\ref{figure: fig/stabilizer_measurement}~(b),
we cycle the ancillary particle $\pa{j}$ around its square using the tunneling operations $X_\x$ and $X_\y$,
while applying nearest-neighbor $\cnot_{\pa{j\pm 1},\pa{j}}$ or $\cnot_{\pa{j},\pa{j\pm 1}}$ immediately before each tunneling.
As defined in Eq.~\eqref{eq: position-dependent spin operation},
the pre- and post-operations $X_\co^{[10,11]}$, $X_\co^{[11,01]}$, or $Z_\co^{[01,10]}$ denote applying $X_\co$ or $Z_\co$ to the two indicated vertices.

From the above discussions, we can partly confirm that the QSD offers the stacking flexibility to the stabilizer measurement, as will be fully confirmed in Sec.~\ref{Section: Scalability Enhancement}.
Namely, our model implements the circuit in Fig.~\ref{figure: fig/stabilizer_measurement}~(b) requiring only nearest-neighbor $\cnot_\co$,
whereas an implementation without QSD would require interactions between particles (qubits) placed at least four qubits apart.
Here, although the present implementation increases the circuit depth compared to the one-particle-one-qubit implementation (i.e., without exploiting QSD),
the increase in failure probability for the stabilizer measurement is expected to be marginal as discussed in Appx.~\ref{Section: Simple Estimate of Failure Probability}.

\newpage
\section{Noise Resilience Enhancement}\label{Section:  Noise Resilience Enhancement}
We show that the GS [Eq.~\eqref{eq: gauge symmetry}] offers resilience
against arbitrary coherent errors or decoherence acting on either spin or position state,
as well as dephasing of both spin and position states (partial loss or complete vanishing of the QSD).
As discussed in the introduction, these would be the dominant noise sources in our architecture, or more generally, in an architecture exploiting QSD.

First, we present a unified noise model containing the above noise sources, and demonstrate that our QEC scheme functions correctly under this noise model.
Our formalization explicitly shows that the lack of the GS renders the unified noise model uncorrectable [see~Eq.~\eqref{eq: gaugesymmetry (derivation of the generalized quantum noise)}],
where the derivation of correctability is inspired by Refs.~\cite{kribs2005unified,nielsen2007algebraic,nielsen2010quantum}.
Second, we explicitly identify and discuss the above three noise sources within this unified noise model.

\subsection{Unified Noise Model}\label{Subsection: Unified Noise Model}

The quantum noise, or more specifically, a nontrivial interaction with the environment ($\E$), that unifies the three types of noise inherent in our architecture, is given by
\begin{align}
    U_E := \sum_{n\in\mathbb{N}_0} E^{(n)} \otimes |e_{n+1}\rangle \langle e_0|_\E,\
    E^{(n)} = \sum_{m\in\{0,1\}^6}\mathfrak{g}_{n,m}\sigma_m,
    \label{eq: unified quantum noise}
\end{align}
where  $|e_i\rangle_\E\in\mathbb{C}^\infty$ is a state of the environment whose initial state is $|e_0\rangle_\E$ defined via purification~\cite{nielsen2010quantum}.
In addition, $\mathfrak{g}_{n,m}\in\mathcal{G}$ is an arbitrary gauge operation chosen such that $\sum_n E^{(n)\dagger} E^{(n)}=I$ (the unitarity condition for $U_E$).

We first demonstrate that our QEC mechanism provides the logical qubit with resilience against this unified quantum noise.
Subsequently, we explicitly validate the Pauli frame update discussed in the previous section under repeated exposure to the quantum noise given by Eq.~\eqref{eq: unified quantum noise}.
Here, to facilitate the two analyses,
we express the recovery operation conditioned on the stabilizer measurement in the Kraus-representation (or operator-sum representation~\cite{nielsen2010quantum}):
\begin{align}
    \mathcal{R}(\rho) = \sum_{m=0}^{2^6-1} \sigma^\dagger_m  P_m^\nodagger \rho P_m^\dagger \sigma_m^\nodagger,\ P_m^\nodagger:= \frac{1}{2^6} \prod_{i=0}^5 \left(I+(-1)^{m_i}s_i\right)
    \label{eq: recovering operation}
\end{align}
where $\rho \in \mathcal{B}((\mathbb{C}^2)^{\otimes 9})$ is a density operator of the three physical particles.
The adjoint action $P_m^\nodagger \cdot P_m^\dagger$ indicates that the outcome $m\in\{0,1\}^6$ is obtained from the stabilizer measurement [cf.~Eq.~\eqref{eq: overall outcome of the stabilizer measurement}],
and $\sigma_m \cdot \sigma^\dagger_m$ is the recovery action corresponding to this outcome.

With the Kraus representation $\mathcal{E}(\rho) := \sum_n E^{(n)} \rho E^{(n)\dagger}$ $(=\tr_\E\!\left(U_\E\,(\rho\otimes|e_0\rangle\langle e_0|_\E)\,U_\E^\dagger\right))$,
we derive the following GS as a demonstration of the resilience against the quantum noise $U_\E$:
\begin{align}
    \mathcal{R}\circ \mathcal{E} \left(|\psi\rangle\langle\psi|\right) \gs |\psi\rangle\langle\psi|,
    \label{eq: gauge symmetry guaranteeing the resilience against unified noise}
\end{align}
where $|\psi\rangle := \alpha |0\rangle_\Lo + \beta |1\rangle_\Lo$ ($\alpha,\beta\in\mathbb{C}$).
This symmetry ensures that the recovery operation removes the influence of the quantum noise on the logical qubit [cf. Fig.~\ref{figure: gauge symmetry}~(b)].
Here is the derivation establishing Eq.~\eqref{eq: gauge symmetry guaranteeing the resilience against unified noise}:
\begin{align}
    |\psi\rangle\langle\psi|
     & \xrightarrow{\mathcal{E}} \sum_{n\in\mathbb{N}_0} \sum_{m\in\{0,1\}^6} \mathfrak{g}_{n,m}^\nodagger \sigma_m^\nodagger |\psi\rangle\langle\psi| \sigma_m^\dagger \mathfrak{g}_{n,m}^\dagger
    \\
     & \gs \sum_{m\in\{0,1\}^6} \sigma_m^\nodagger|\psi\rangle\langle\psi| \sigma_m^\dagger
    \label{eq: gaugesymmetry (derivation of the generalized quantum noise)}                                                                                                                                                    \\
     & \xrightarrow{\mathcal{R}} \sum_{m,m^\prime\in\{0,1\}^6} \sigma_{m^\prime}^\dagger P_{m^\prime}^\nodagger \sigma_m^\nodagger  |\psi\rangle\langle\psi| \sigma_m^\dagger P_{m^\prime}^\dagger \sigma_{m^\prime}^\nodagger
    \\
     & = \sum_{m\in\{0,1\}^6}  \sigma_{m}^\dagger \sigma_{m}^\nodagger |\psi\rangle\langle\psi|\sigma_{m}^\nodagger\sigma_{m}^\dagger.
    \label{eq: derivation of the generalized quantum noise}
\end{align}
The normalization constant is omitted in the second line, and thus the final state reduces to $|\psi\rangle\langle\psi|$.
Here, we used the GS~[Eq.~\eqref{eq: gauge symmetry}] in the second step,
and two identities in the final step:
$P_{m^\prime}^\nodagger = \sigma_{m^\prime}^\nodagger P_0 \sigma^\dagger_{m^\prime}$ and $P_0 \sigma^\dagger_{m^\prime}\sigma_m^\nodagger |\psi\rangle = \delta_{m^\prime,m}|\psi\rangle$ for all $m,m^\prime \in \{0,1\}^6$
(we retain $\sigma_{m}^\dagger \sigma_{m}^\nodagger (=I)$ for later convenience in deriving Eq.~\eqref{eq: pauli frame update}).

We further obtain a symmetry that validates the Pauli-Frame update discussed in Sec.~\ref{Subsection: Stabilizer measurement},
where the recovery operations are collectively applied according to the outcomes of $N$ rounds of stabilizer measurements:
\begin{align}
    (\mathcal{R}\circ\mathcal{E})^N (|\psi\rangle \langle\psi|) \gs \sum^{2^{6N}-1}_{M=0} \left(\bigcirc^{N-1}_{i=0} \mathrm{Ad}(\sigma^\dagger_{m^i})\right)\left(\bigcirc^{N-1}_{i=0} \mathrm{Ad}(\sigma_{m^i}^\nodagger)\right) (|\psi\rangle \langle\psi|),
    \label{eq: pauli frame update}
\end{align}
where $m^{i}\in\{0,1\}^6$ and $M=\sum^{N-1}_{i=0}2^{6i}m^i$.
Here, we denote the adjoint action as $\mathrm{Ad}(\sigma_{m^i}^\nodagger)(\rho):=\sigma_{m^i}^\nodagger \rho \sigma^\dagger_{m^i}$,
and the composition as $\bigcirc^{N-1}_{i=0} \mathrm{Ad}(\sigma_{m^i}) := \mathrm{Ad}(\sigma_{m^{N-1}})\circ \mathrm{Ad}(\sigma_{m^{N-2}})\circ \cdots \circ \mathrm{Ad}(\sigma_{m^0}$).
This follows from repeated application of Eq.~\eqref{eq: derivation of the generalized quantum noise}. For instance, in the case of $N=2$, we have:
\begin{align}
    (\mathcal{R}\circ \mathcal{E})^2 \left(|\psi\rangle\langle\psi|\right)
     & \gs (\mathcal{R}\circ \mathcal{E}) \left(\sum_{m^0\in\{0,1\}^6} \sigma_{m^0}^\dagger \sigma_{m^0}^\nodagger |\psi\rangle\langle\psi|\sigma_{m^0}^\dagger \sigma_{m^0}^\nodagger\right)
    \\
     & \gs  \sum_{m^0,m^1\in\{0,1\}^6} \sigma_{m^1}^\dagger\sigma_{m^0}^\dagger \sigma_{m^1}^\nodagger \sigma_{m^0}^\nodagger |\psi\rangle\langle\psi| \sigma_{m^0}^\dagger \sigma_{m^1}^\dagger \sigma_{m^0}^\nodagger \sigma_{m^1}^\nodagger,
    \label{eq: }
\end{align}
where we used $\sigma^\dagger_{m^i} \sigma^\nodagger_{m^j} = \pm \sigma^\nodagger_{m^j} \sigma_{m^i}^\dagger$ ($i,j\in\mathbb{N}_0$) in the last step.

\subsection{Three Types of Dominant Noise}\label{Subappendix: Three Types of Dominant Noise}
We then explicitly identify and discuss the three types of dominant noise within the unified correctable noise [Eq.~\eqref{eq: unified quantum noise}].
The fact that our scheme achieves resilience against these noise sources implies that exploiting the GS
is an essential approach for introducing a QEC mechanism into an architecture that utilizes the QSD, not limited to the present architecture.

\paragraph{Unified Spin-related Noise}
We first identify a noise model acting on the spin state of a particle,
which provides a unified description of both arbitrary coherent errors and decoherence.
A key feature of this model is that its action varies depending on the particle's position, reflecting realistic conditions in our QSD-based architecture.

Given that the projector onto an arbitrary position state can be written as $|n_1,n_0\rangle\langle n_1, n_0|_\po$ $=\frac{1}{4}(I+(-1)^{n_1}Z)_\x(I+(-1)^{n_0}Z)_\y$ for $n_1,n_0\in\{0,1\}$,
the unified spin-related noise is explicitly written as follows (letting $n_0$ and $n_1$ be the 0th and 1st digits of $n$, respectively):
\begin{align}
    (U_{E,\co})_\pa{i} & = \sum_{n\in\mathbb{N}_0} \left(E^{(n)}_\co\right)_\pa{i}\otimes |e_{n+1}\rangle\langle e_0|_E,
    \label{eq: unified spin-related noise}
    \\
    E_\co^{(n)}       & := M_\co^{(n)}(I_\x+(-1)^{n_1}Z_\x)(I_\y+(-1)^{n_0}Z_\y).
    \label{eq: element of spin-related noise}
\end{align}
where the actions on the spin-state $\{M_\co^{(n)}\in\mathrm{End}(\mathbb{C}^2)\}$ are chosen so that $\sum_nE_\co^{(n)\dagger}E_\co^{(n)}=I$,
and we label these actions by $n$.
Given that $M_\co^{(n)}$ belongs to an algebra generated by $I_\co$, $X_\co$, and $Z_\co$,
the operator element $E^{(n)}_\co$ is clearly captured by the unified operator $E^{(n)}$ in Eq.~\eqref{eq: unified quantum noise}.

One of the dominant errors or decoherence contained in Eq.~\eqref{eq: unified spin-related noise} may be amplitude damping (or energy dissipation) of the spin state.
As an extreme case, when $M_\co^{(n)}$ takes $|0\rangle\langle 0|_\co$ for $0\leq n\leq 3$ and $|0\rangle\langle 1|_\co$ for $4\leq n\leq 7$, a particle affected by this noise deterministically relaxes its spin state to only $|0\rangle_\co$.

\paragraph{Unified Position-related Noise}
From Eq.~\eqref{Subsection: Unified Noise Model}, we subsequently identify a unified noise model acting on the position state of a particle.
As discussed below, this model serves to describe the failure of the gate operation acting on the particle's position state.

Analogously to the unified spin-related noise in Eq.~\eqref{eq: unified spin-related noise},
the unified position-related noise is given by
\begin{align}
    (U_{E,\po})_\pa{i} & = \sum_n \left(\delta_{n_2, 0} E_\x^{(n)}+\delta_{n_2, 1} E_\y^{(n)}\right)_\pa{i}\otimes |e_{n+1}\rangle\langle e_0|_E,
    \label{eq: unified position-related noise}
    \\
    E_\x^{(n)}        & := (I_\co+(-1)^{n_1}Z_\co)M_\x^{(n)}(I_\y+(-1)^{n_0}Z_\y),
    \\
    E_\y^{(n)}        & := (I_\co+(-1)^{n_1}Z_\co)(I_\x+(-1)^{n_0}Z_\x)M_\y^{(n)},
\end{align}
where $n_i\in\{0,1\}$ is the $i$th digit of $n$ for $0 \leq i\leq 2$.
In addition, $M_\x^{(n)}$ (resp.~$M_\y^{(n)}$) is in an algebra generated by $I_\x$, $Z_\x$, and $X_\x$ (resp.~$I_\y$, $Z_\y$, and $X_\y$).

In this unified noise model, no operator elements ($E_\x^{(n)}$ and $E_\y^{(n)}$) include the Pauli-X operations acting on both $x$- and $y$-axis position states
(which are uncorrectable within our QEC).
This setting is physically more reasonable than one including an operator element acting on both states.
Namely, under our assumption that a single particle resides at the square vertices with no diagonal connections,
such undesired Pauli $X$ operations would occur only through two successive noise events rather than a single one.

One of the dominant noise sources in our architecture, which falls within Eq.~\eqref{eq: unified position-related noise},
is the failure of a gate operation acting on the particle's position state.
As an example, when $E^{(n)}_\x$ alternately takes $|1\rangle\langle 0|_\x$ and $|1\rangle\langle 1|_\x$ while $E_y^{(n)}$ takes a zero matrix,
the particle affected by this noise deterministically localizes in a region corresponding to the state $|1\rangle_\x$.
This situation would physically represent the phenomenon in which the particle succeeds in tunneling ($X_\x$ or $X_\y$) from one region to another $(|0\rangle_\x \rightarrow |1\rangle_\x)$
while failing to tunnel in the opposite direction $(|1\rangle_\x \rightarrow |0\rangle_\x)$.

\paragraph{Dephasing noise (loss of QSD)}
Finally, we discuss dephasing noise acting on both spin and position states.
Such environmental noise degrades the QSD of a particle, resulting in a mixed state.
As shown below, the unified correctable noise model in Eq.~\eqref{eq: unified quantum noise} contains a dephasing under which
a particle reduces to a classical state, i.e., its QSD completely vanishes.

The most physically natural noise of dephasing contained in Eq.~\eqref{eq: unified quantum noise} is the following model:
\begin{align}
     & \left(U_{E, \mathrm{loss}}\right)_\pa{j} := \sum_{n\in\{0,1\}^3} \left(E^{(n)}_{\mathrm{loss}}\right)_\pa{j} \otimes |e_{n+1}\rangle \langle e_0|_E,
    \label{eq: dephasing noise}
    \\
     & E^{(n)}_{\mathrm{loss}}:= \alpha_n\left[I_\co I_\x I_\y - \frac{1}{8}\bigl(I_\co + (-1)^{n_2}Z_\co\bigr)\bigl(I_\x + (-1)^{n_1}Z_\x\bigr)\bigl(I_\y + (-1)^{n_0}Z_\y\bigr)\right].
    \label{eq: operator element (dephasing noise) loss}
\end{align}
where $\alpha_n\in\mathbb{C}$ is chosen so that $\sum_n E^{(n)\dagger}_{\mathrm{loss}} E^{(n)}_{\mathrm{loss}}=I$.
Given that the second term in Eq.~\eqref{eq: operator element (dephasing noise) loss} can be written as $|n_2, n_1, n_0\rangle\langle n_2, n_1, n_0|_{\co,\po}$,
each noise event removes one set of position and spin states from the QSD of the particle.
If the probabilities of such losses are equal, the coefficients satisfy $|\alpha_n|^2 = 1/7$ for all $n$ ($=n_2 n_1 n_0$).

We note that, as an extreme case, the unified correctable noise model~[Eq.~\eqref{eq: unified quantum noise}] even contains the following dephasing noise
which causes the QSD of a particle to fully vanish, resulting in a classical state upon a single application:
\begin{align}
     & \left(U_{E, \mathrm{vanish}}\right)_\pa{j} := \sum_{n\in\{0,1\}^3} \left(E^{(n)}_{\mathrm{vanish}}\right)_\pa{j} \otimes |e_{n+1}\rangle \langle e_0|_E,
    \label{eq: dephasing noise (vanish)}
    \\
     & E^{(n)}_{\mathrm{vanish}}:= \frac{1}{8}\bigl(I_\co + (-1)^{n_2}Z_\co\bigr)\bigl(I_\x + (-1)^{n_1}Z_\x\bigr)\bigl(I_\y + (-1)^{n_0}Z_\y\bigr).
    \label{eq: operator element (dephasing noise) vanish}
\end{align}
Seven or more applications of $\left(U_{E, \mathrm{loss}}\right)_\pa{j}$ yield the same operator elements as those in Eq.~\eqref{eq: operator element (dephasing noise) vanish}.

\begin{figure}[t]
    \centering
    \begin{minipage}[b]{0.39\columnwidth}
        \centering
        \includegraphics[width=5cm]{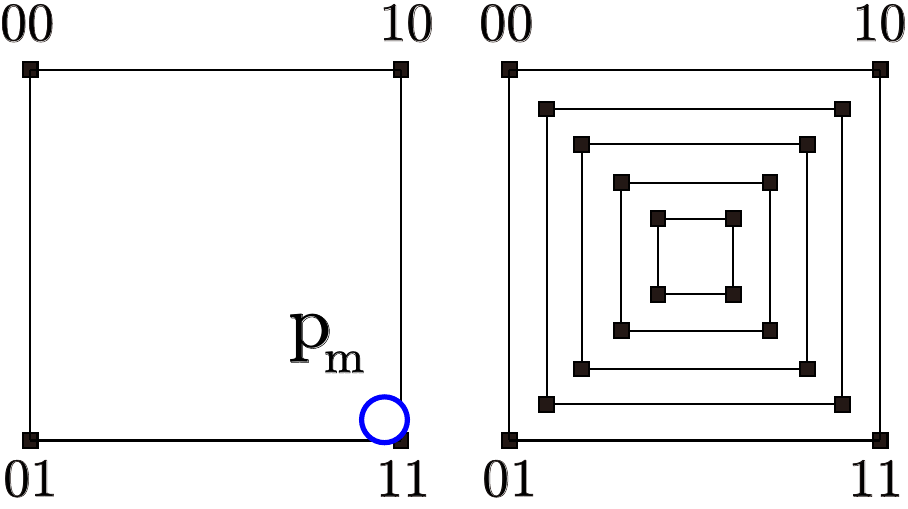}
        \subcaption{}
        \label{figure: }
    \end{minipage}
    \begin{minipage}[b]{0.59\columnwidth}
        \centering
        \includegraphics[width=8.5cm]{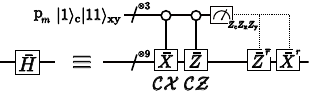}
        \subcaption{}
        \label{figure: hadamard circuit}
    \end{minipage}
    \caption{
    Setup for the logical Hadamard gate that converts the $\bar{Z}$ eigenstates of the three-particle system to $\bar{X}$ eigenstates, and vice versa.
    (a) A square in which a particle ($\pa{m}$; mediator particle) resides and the nested-square system to which the logical Hadamard gate is applied.
    This additional square can be adjacent to the system either horizontally or vertically.
    Furthermore, the particle $\pa{m}$ can take any state within the $-1$ eigenspace of $Z_\co Z_\x Z_\y$.
    (b) Gate-teleportation circuit for the logical Hadamard gate.
    The meter symbol on the upper line represents a measurement of $Z_\co Z_\x Z_\y$ for the particle $\pa{m}$,
    and its result is denoted by $(-1)^{r}$ ($r\in\{0,1\}$).
    $\bar{Z}^{\bar{r}}$ (resp.~$\bar{X}^{r}$) indicates that 
    $\bar{Z}$ (resp.~$\bar{X}$) is applied to the system depending on the result $\bar{r}$ (resp.~$r$), where $\bar{r} := 1 - r$.
    }
\end{figure}

\section{Scalability Enhancement}\label{Section: Scalability Enhancement}
In this section, we show that the integrated use of the QSD and the GS offers a form of architectural flexibility, i.e., stacking flexibility, for quantum computation within quantum error correction.
Specifically, we discuss how the three quantum gates ($\cx$, $\sqrt{\cx}$, and $\cz$) can be implemented in a manner applicable to two systems aligned either horizontally or vertically,
while the resource efficiency, the inherent advantage of the nested-square system, is preserved.

These three gates together realize the logical Hadamard and Toffoli gates in our QEC system and further enable a space-efficient realization of a quantum adder (see Fig.~\ref{figure: adder}).
Given that the universality of the Hadamard and Toffoli gates has been proven~\cite{shi2002both,aharonov2003simple}, 
combining the insights from the present and the companion papers~\cite{asaka2025quantum} would pave the way toward highly scalable ultrafast fault-tolerant quantum computation.

The key to the stacking flexibility is the following GS, which localizes the logical operator $\bar{X}$ onto the outermost particle:
\begin{align}
    \bar{X} = (X_\co X_\x X_\y)_{\pa{4}} s_4 \propto (X_\co X_\x X_\y)_{\pa{4}}.
    \label{eq: gauge symmetry (logical X)}
\end{align}
Here, we assume that the $s_0$--$s_5$ eigenvalues are all $+1$, as restored by the recovery operation [Eq.~\eqref{eq: recovery operation}].
Otherwise, for the $\cx$ and the $\cz$ gates, the eigenvalue-dependent Pauli byproducts can be tracked from the history of the stabilizer measurements
and compensated by updating the recovery operation (Pauli frame) in Eq.~\eqref{eq: pauli frame update} accordingly.
In contrast, the byproduct for $\sqrt{\cx}$ is non-Pauli, and hence we must restore the $s_4$ eigenvalues of both systems to $+1$
by a simple Pauli $Z$-operation, with the Pauli frame updated accordingly.

\subsection{$\cx$ gate}\label{Subsection: cx gate}
The $\cx$ gate applies the logical operator $\bar{X}$ to a target three-particle system $\{\pa{4},\pa{2},\pa{0}\}$ ($\tgt$)
if the control system $\{\paprime{4},\paprime{2},\paprime{0}\}$ ($\ctrl$) has eigenvalue $-1$ under $\bar{X}$.
Explicitly, it is defined as
\begin{align}
    \cx_{\ctrl, \tgt}         :  = \frac{1}{2}(\bar{I}+\bar{X})_\ctrl\otimes \bar{I}_\tgt + \frac{1}{2}(\bar{I}-\bar{X})_\ctrl\otimes \bar{X}_\tgt.
    \label{eq: CX definition}
\end{align}
As will be discussed below, this gate can be implemented using the same scheme regardless of whether the two systems are aligned horizontally or vertically
by requiring nearest-neighbor interactions only between the outermost particles $\pa{4}$ and $\paprime{4}$.

Through communication between only the outermost particles
of the two systems, specifically, $(\cnot_\co)_{\pa{4},\paprime{4}}$, we implement $\cx$ gate.
The implementation relies on the following symmetry:
\begin{align}
    \cx_{\ctrl,\tgt} & \propto \sum_{p=0}^1 \left(\frac{I_\co I_\x I_\y + (-1)^p X_\co X_\x X_\y}{2}\right)_{\paprime{4}} (X_\co X_\x X_\y)^p_\pa{4},
    \label{eq: derivation CX (first step)}
\end{align}
where we use the symmetry in Eq.~\eqref{eq: gauge symmetry (logical X)} to both the control and target systems.
By further transforming the RHS, we derive the specific implementation scheme as
\begin{align}
                     & =(H_\co H_\x H_\y)^{\otimes 2}_{\paprime{4},\pa{4}}\left(X_\co^{[01,10]}\right)_\paprime{4}
    \notag                                                                                                                                                                                                                                         \\
                     & \qquad\quad\left(\sum_{p=0}^{1} \left(|p\rangle\langle p|_\co\right)_\paprime{4}(Z_\co Z_\x Z_\y)_\pa{4}^p\right)\left(X_\co^{[01,10]}\right)_\paprime{4}(H_\co H_\x H_\y)^{\otimes 2}_{\paprime{4},\pa{4}}
    \label{eq: derivation CX (second step)}\\
                     & =(H_\co H_\x H_\y)^{\otimes 2}_{\paprime{4},\pa{4}}\left(X_\co^{[01,10]}\right)_{\paprime{4},\pa{4}}^{\otimes 2}
    \notag                                                                                                                                                                                                                                         \\
                     & \qquad\quad \left(\sum_{p=0}^{1} (Z_\co)_\paprime{4}^p \left(|p\rangle\langle p|_\co\right)_\pa{4}\right)\left(X_\co^{[01,10]}\right)_{\paprime{4},\pa{4}}^{\otimes 2}(H_\co H_\x H_\y)^{\otimes 2}_{\paprime{4},\pa{4}}
    \\
                     & =\left(H_\x H_\y\cdot Z_\co^{[01,10]} \right)_\paprime{4} \left(H_\co H_\x H_\y\cdot X_\co^{[01,10]}\right)_\pa{4}
    \notag                                                                                                                                                                                                                                         \\
                     & \qquad\quad \left(\sum_{p=0}^{1} (X_\co)_\paprime{4}^p \left(|p\rangle\langle p|_\co\right)_\pa{4}\right)\left(Z_\co^{[01,10]}\cdot H_\x H_\y\right)_\paprime{4} \left(X_\co^{[01,10]}\cdot H_\co H_\x H_\y \right)_\pa{4}.
    \label{eq: derivation CX (final step)}
\end{align}
Namely, to implement $\cx$, we apply $X_\co$ to $\paprime{4}$ conditioned on the spin state of $\pa{4}$ being $|1\rangle_\co$,
together with the applications of $Z_\co^{[01,10]} \cdot H_\x H_\y$ (and~$H_\x H_\y\cdot Z_\co^{[01,10]}$) and $X_\co^{[01,10]}\cdot H_\co H_\x H_\y$ (and~$H_\co H_\x H_\y\cdot X_\co^{[01,10]}$) to $\paprime{4}$ and $\pa{4}$, respectively, before (and~after) this conditional spin flip.
Here, the superscript $[01,10]$ denotes position dependence as defined in Eq.~\eqref{eq: position-dependent spin operation}.
Note that we describe the concrete procedure for this conditional spin flip two paragraphs below.

Here, the above derivation [Eqs.~\eqref{eq: derivation CX (second step)}--\eqref{eq: derivation CX (final step)}] relies on the following key relations.
In the first step, we have applied the following identity to $\paprime{4}$:
\begin{align}
    \frac{I_\co I_\x I_\y + (-1)^p Z_\co Z_\x Z_\y}{2}
     & = (|p00\rangle\langle p00| + |\bar{p}10\rangle\langle \bar{p}10| + |p11\rangle\langle p11| + |\bar{p}01\rangle\langle \bar{p}01|)_{\co,\po}
    \label{eq: derivation CX (key relation: second step (1))}
    \\
     & = X_\co^{[01,10]} |p\rangle\langle p|_\co X_\co^{[01,10]}
    \label{eq: derivation CX (key relation: second step (2))}
\end{align}
where $\bar{p}:=1-p$, and $|cxy\rangle_{\co,\po}:=|c\rangle_\co |xy\rangle_\po$ for $c,x,y\in\{0,1\}$.
In the second step, we have first used the identity
\begin{align}
    \sum^1_{p=0} (|p\rangle\langle p|_\co)_\paprime{4} (Z_\co Z_\x Z_\y)^p_\pa{4}
     & =\sum^1_{p=0} \frac{1}{2}(I_\co + (-1)^p Z_\co)_\paprime{4} (Z_\co Z_\x Z_\y)^p_\pa{4}
    \\
     & = \sum^1_{p=0}(Z_\co)^p_\paprime{4} \left(\frac{I_\co I_\x I_\y + (-1)^p Z_\co Z_\x Z_\y}{2}\right)_\pa{4},
    \label{q: derivation CX (key relation: third step (2))}
\end{align}
and then also applied the identity given in Eqs.~\eqref{eq: derivation CX (key relation: second step (1))}--\eqref{eq: derivation CX (key relation: second step (2))} to $\pa{4}$.
In the final step, we have used the relations $H_\co X_\co^{[01,10]} = Z_\co^{[01,10]}H_\co$ and $H_\co Z_\co^p H_\co = X_\co^p$ for $\paprime{4}$.

To apply the conditional spin flip $\sum_{p=0}^{1} (X_\co)_\paprime{4}^p \left(|p\rangle\langle p|_\co\right)_\pa{4}$ in Eq.~\eqref{eq: derivation CX (final step)}, i.e., to flip the spin state of $\paprime{4}$ conditioned on the spin state of $\pa{4}$,
we sequentially apply $(X_\y)_\pa{4}$, $(X_\x)_\paprime{4}$ and $(X_\y)_\pa{4}$
with a $\cnot_{\pa{4},\paprime{4}}$ inserted immediately before each tunneling operation ($X_\x$ or $X_\y$).
When the two systems are aligned horizontally, this procedure can be illustrated as follows:
\begin{align}
    \begin{array}{c}
        \includegraphics[height=2.7cm]{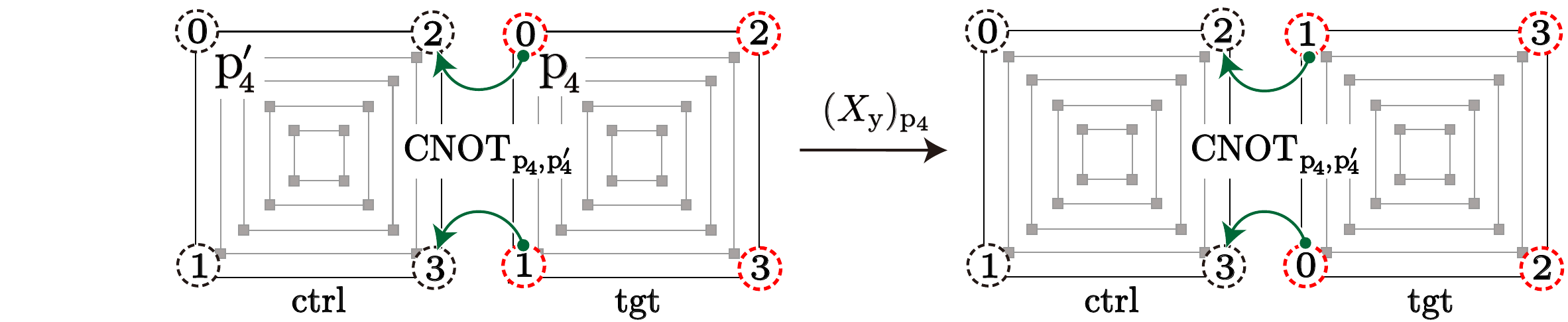},
    \end{array}
    \label{eq: fig/CX_1.pdf}
    \\
    \begin{array}{c}
        \includegraphics[height=2.7cm]{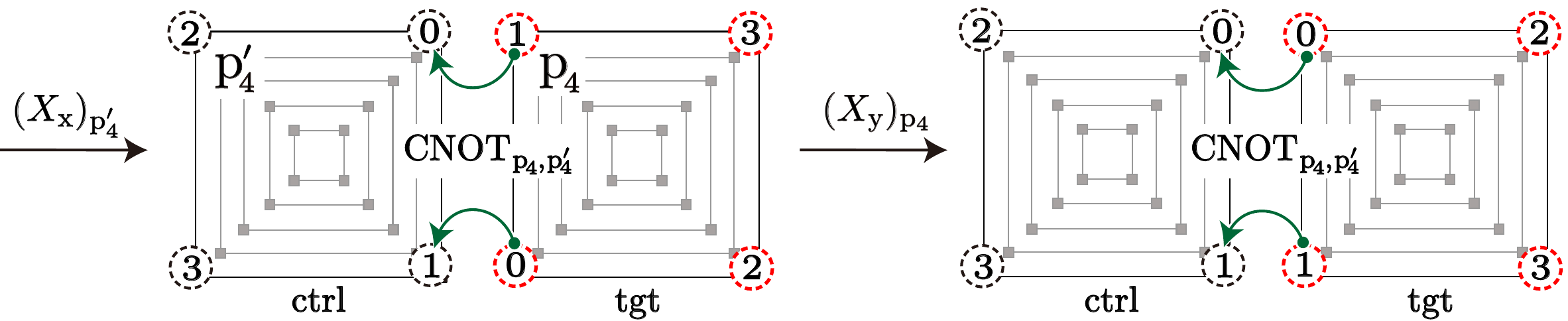}.
    \end{array}
    \label{eq: fig/CX_2.pdf}
\end{align}
where the red circles represent states of $\pa{4}$ with the spin $|1\rangle_\co$,
and the black circles represent states of $\paprime{4}$ with an arbitrary spin state.
Here, the numbers ($0$--$3$) inside the circles denote the initial position state ($|00\rangle_\po$--$|11\rangle_\po$) before the procedure begins.

When the systems are aligned vertically, the aforementioned procedure completes the conditional spin flip
since the nearest-neighbor $(\cnot_\co)_{\pa{4},\paprime{4}}$ can be applied at every vertex.
Meanwhile, for horizontally aligned systems, we need an additional step to apply the $\cnot_{\pa{4},\paprime{4}}$ to the remaining states of the target system,
i.e., the red circles numbered $2$ and $3$ in Eqs.~\eqref{eq: fig/CX_1.pdf} and \eqref{eq: fig/CX_2.pdf}.
To this end, we apply the tunneling operation $(X_\x)_\pa{4}$
and repeat the procedures in Eqs.~\eqref{eq: fig/CX_1.pdf} and \eqref{eq: fig/CX_2.pdf}.
Here, we treat the residual $(X_\x)_\pa{4}$  (or $(X_\x)_\paprime{4}$ when the systems are aligned vertically) as the detected error by updating the Pauli frame [cf. Eq.~\eqref{eq: pauli frame update}].

\subsection{$\sqrt{\cx}$ gate}\label{Subsection: root cx gate}
We hereafter refer to the following conditional operations as the $\sqrt{\cx}$ gate:
\begin{align}
    \sqrt{\cx}_{\ctrl, \tgt}  :  = & \frac{1}{2}(\bar{I}+\bar{X})_\ctrl\otimes \bar{I}_\tgt + \frac{1}{2}(\bar{I}-\bar{X})_\ctrl\otimes \sqrt{\bar{X}}_\tgt.
    \label{eq: CXroot definition}
    \\
    \sqrt{\bar{X}}_\tgt:=          & (V_\co V_\x V_\y)_\pa{4},\ V:=\frac{1+i}{2}\left(\begin{matrix}1&-i\\-i&1\end{matrix}\right).
    \label{eq: root X definition}
\end{align}
Here, we obtain the identity $(\sqrt{\bar{X}})^2 = (X_\co X_\x X_\y)_\pa{4} \propto \bar{X}$ and hence $(\sqrt{\cx})^2 \propto \cx$, which underlies the implementation of the Toffoli gate as shown in Fig.~\ref{figure: adder}~(a).
We note that this gate becomes practical only when it is applied twice.
Namely, the $s_0$--$s_5$ stabilizer measurement loses its effectiveness if we apply $\sqrt{\mathcal{CX}}$ only once,
since $\sqrt{\bar{X}}$ maps the stabilizer state to a different simultaneous eigenspace stabilized by $\langle \sqrt{\bar{X}} s_i \sqrt{\bar{X}}^\dagger \mid 0\leq i \leq 5\rangle$.

We implement the $\sqrt{\cx}_{\ctrl, \tgt}$ gate by a similar procedure to that for the $\cx$ gate.
Namely, the implementation is achieved via communication between the two outermost particles $\paprime{4}$ and $\pa{4}$,
or more specifically, $\cs_{\paprime{4}, \pa{4}}$, $\cpi_{\paprime{4}, \pa{4}}$, and $\cpihalf_{\paprime{4}, \pa{4}}$ gates.
A concrete procedure is derived as follows:
\begin{align}
    \sqrt{\cx}_{\ctrl, \tgt}
    \propto & \sum_{p=0}^1  \left(\frac{I_\co I_\x I_\y + (-1)^p X_\co X_\x X_\y}{2}\right)_{\paprime{4}} (V_\co V_\x V_\y)_\pa{4}^p
    \label{derivation rootcx (first step)}                                                                                     \\
    = & \left(H_\co H_\x H_\y\cdot X_\co^{[01,10]}\right)_\paprime{4}\left(H_\co H_\x H_\y\right)_\pa{4}
    \\
      & \quad \left(\sum_{p=0}^1 \left(|p\rangle\langle p|_\co\right)_\paprime{4} (S_\co S_\x S_\y)_\pa{4}^p\right)
    \left(X_\co^{[01,10]}\cdot H_\co H_\x H_\y\right)_\paprime{4}\left(H_\co H_\x H_\y\right)_\pa{4}
    \label{derivation rootcx (final step)}
\end{align}
where $S:= |0\rangle \langle 0| + i |1\rangle\langle 1|$.
The derivation is based on Eq.~\eqref{eq: gauge symmetry (logical X)} for the control system, and Eqs.~\eqref{eq: derivation CX (key relation: second step (1))} and \eqref{eq: derivation CX (key relation: second step (2))}.

We effectively apply the conditional operation $\sum_{p=0}^1 \left(|p\rangle\langle p|_\co\right)_\paprime{4} (S_\co S_\x S_\y)_\pa{4}^p$ using the identity $S_\co S_\x S_\y = i^{x+y} S_\co$,
where $x,y\in\{0,1\}$ denote the x- and y-axis position states of the target particle, i.e., $\pa{4}$.
Namely, we shift the position of $\paprime{4}$ by $(X_\x \rightarrow X_\y \rightarrow X_\x)^2$ (resp.~$X_\x \rightarrow X_\y \rightarrow X_\x$) if the two systems are aligned horizontally (resp.~vertically),
and apply $\cs_{\paprime{4},\pa{4}}$, $\cs_{\paprime{4},\pa{4}}\cpihalf_{\paprime{4},\pa{4}}$, and $\cs_{\paprime{4},\pa{4}}\cpi_{\paprime{4},\pa{4}}$
to $\pa{4}$ with position state $|00\rangle_\po$, $|01\rangle_\po$ (and $|10\rangle_\po$), and $|11\rangle_\po$, respectively, before each tunneling operation
(these interactions are defined in Eqs.~\eqref{eq: definition of nearest-neighbor cs}--\eqref{eq: definition of nearest-neighbor cshalf}).
This procedure can be verified using the following identities:
\begin{align}
    (\mathrm{C}\theta)_{\paprime{4},\pa{4}} =
    \sum_{xy}\sum_{p=0}^1 \left(|p\rangle\langle p|_\co\otimes|xy\rangle\langle xy|_\po\right)_\paprime{4} ((e^{i\theta} I_\co)^p \otimes|xy\rangle\langle xy|_\po)_\pa{4},
    \label{eq: other form of cpi and cpihalf}
\end{align}
for $\theta\in\{\pi/2, \pi\}$.

The above procedure for the two systems aligned horizontally can be illustrated as follows:
\begin{align}
    \begin{array}{c}
        \includegraphics[height=2.7cm]{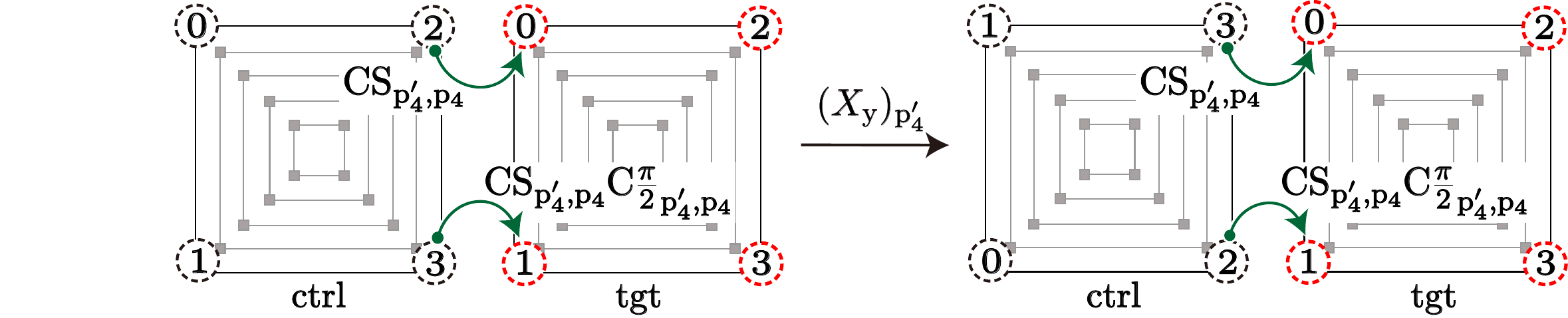},
    \end{array}
    \label{eq: fig/rootCX1.pdf}
    \\
    \begin{array}{c}
        \includegraphics[height=2.7cm]{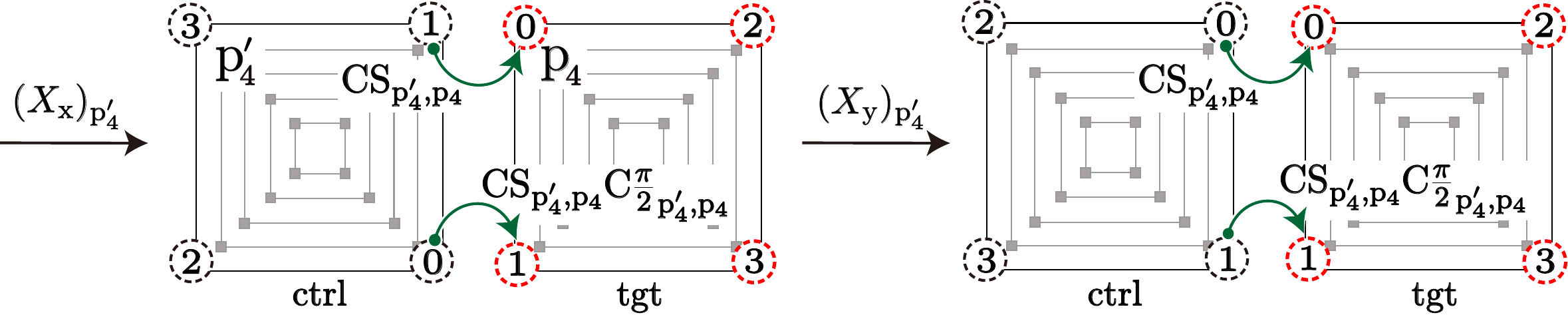}.
    \end{array}
    \label{eq: fig/rootCX2.pdf}
\end{align}
In addition, we apply $\cs_{\paprime{4},\pa{4}}\cpihalf_{\paprime{4},\pa{4}}$ (resp.~$\cs_{\paprime{4},\pa{4}}\cpi_{\paprime{4},\pa{4}}$) to vertices labeled 10 and 00 (resp.~11 and 01)
of the control and target systems, and then move $\pa{4}$ in the same way as the above illustrations.

We can slightly reduce the implementation overhead for the horizontally aligned systems by
replacing the operator $(H_\y)_\pa{4}$ applied before and after the conditional operation in Eq.~\eqref{derivation rootcx (final step)} with $X_\y H_\y$ and $H_\y X_\y$, respectively.
With this, the conditional operation we need to implement is changed to
\begin{align}
    \sum_{p=0}^1 \left(|p\rangle\langle p|_\co\right)_\paprime{4} (S_\co S_\x (X_\y S_\y X_\y))_\pa{4}^p.
\end{align}
Thereby, we only need to apply the same gate $\cs_{\paprime{4},\pa{4}}\cpihalf_{\paprime{4},\pa{4}}$
to the particles located on the vertices labeled $11$ and $01$, instead of converting this gate to $\cs_{\paprime{4},\pa{4}}\cpi_{\paprime{4},\pa{4}}$ as described below Eq.~\eqref{eq: fig/rootCX2.pdf}.
The operator $\cs_{\paprime{4},\pa{4}}$ in Eqs.~\eqref{eq: fig/rootCX1.pdf}--\eqref{eq: fig/rootCX2.pdf} is changed to $\cs_{\paprime{4},\pa{4}}\cpi_{\paprime{4},\pa{4}}$.

\subsection{$\cz$ gate}\label{Subsection: cz gate}
The $\cz$ gate applies the logical $\bar{Z}$ to the target system
conditioned on whether the $\bar{X}$ eigenvalue of the control system is $-1$.
Explicitly, it is defined by
\begin{align}
    \cz_{\ctrl, \tgt}
    := & \frac{1}{2}(\bar{I}+\bar{X})_\ctrl\otimes \bar{I}_\tgt + \frac{1}{2}(\bar{I}-\bar{X})_\ctrl\otimes \bar{Z}_\tgt.
    \label{eq: CZ definition}
\end{align}
This gate is identical to the logical CNOT gate as will be seen in Eq.~\eqref{eq: equivalence between CZ and CNOT}.
We achieve the implementation via pairwise communication among the three pairs of particles $(\paprime{2i}, \pa{2i})$ for $0\leq i \leq 2$, i.e., $(\cnot_\co)_{\pa{2i},\paprime{2i}}$,
when the two systems are aligned vertically.
For the two horizontally aligned systems, we execute $\cz$ using the next-nearest-neighbor $(\cnot_\co)_{\pa{1},\pa{3}}$ and $(\cnot_\co)_{\pa{3},\paprime{4}}$.

\paragraph{Vertical}
When the two systems are aligned vertically, we rely on the following identity for implementing $\cz$:
\begin{align}
    \cz_{\ctrl,\tgt}
    =   \bar{I}_\ctrl\otimes\frac{1}{2} (\bar{I}+\bar{Z})_\tgt + \bar{X}_\ctrl\otimes \frac{1}{2}(\bar{I}-\bar{Z})_\tgt.
    \label{eq: equivalence between CZ and CNOT}
\end{align}
Specifically, we effectively apply this gate through the following scheme
(see the next paragraph for justification of this):
\begin{align}
     & \bigotimes^2_{i=0}\left(\sum_{p=0}^1 (X_\co X_\x X_\y)_\paprime{2i}\left(\frac{I_\co I_\x I_\y + (-1)^p Z_\co Z_\x Z_\y}{2}\right)_{\pa{2i}}\right)
    \label{eq: gauge symmetry of CZ (vertical)}
    \\
     & =    \bigotimes^2_{i=0}  \Biggl(\left(H_\x H_\y\cdot Z_\co^{[01,10]}\right)_\paprime{2i} \left(X_\co^{[01,10]}\right)_\pa{2i}\Biggr.
    \\
     & \Biggl.\qquad\qquad \Bigl(\sum_{p=0}^1 (X_\co)^p_\paprime{2i} (|p\rangle\langle p|_\co)_\pa{2i}\Bigr) \left(Z_\co^{[01,10]}\cdot H_\x H_\y\right)_\paprime{2i} \left(X_\co^{[01,10]}\right)_\pa{2i}\Biggr),
\end{align}
where, in the second step, we use some relations presented in the paragraph around Eqs.~\eqref{eq: derivation CX (key relation: second step (1))}--\eqref{q: derivation CX (key relation: third step (2))}.
Namely, we apply the conditional operation $(X_\co)^p_\paprime{2i} (|p\rangle\langle p|_\co)_\pa{2i}$,
along with $Z_\co^{[01,10]}\cdot H_\x H_\y$ (and~$H_\x H_\y\cdot Z_\co^{[01,10]}$) and $X_\co^{[01,10]}$ applied to $\paprime{2i}$ and $\pa{2i}$ ($\forall 0\leq i\leq2$), respectively, both before (and~after) this conditional operation.
The scheme for this conditional operation is the same as that described around Eqs.~\eqref{eq: fig/CX_1.pdf}--\eqref{eq: fig/CX_2.pdf}.

Based on two observations as follows, we verify that the scheme given by Eq.~\eqref{eq: gauge symmetry of CZ (vertical)} is effectively identical to $\cz$ [Eq.~\eqref{eq: equivalence between CZ and CNOT}].
First, when the $\bar{Z}_\tgt$ eigenvalue is negated, an odd number of particles $\{\pa{2i}\}$ are in the $-1$ eigenstate under $(Z_\co Z_\x Z_\y)_{\pa{2i}}$,
which follows from $\bar{Z}_\tgt:= \bigotimes^2_{i=0}(Z_\co Z_\x Z_y)_{\pa{2i}}$.
Second, when an odd (resp.~even) number of operators $X_\co X_\x X_y$ act on the particles $\{\pa{2i}\}$,
the control system is affected by one of the following four operations:
\begin{align}
    \bar{X},\ \bar{X}s_4,\ \bar{X}s_4 s_5, \bar{X} s_5\
    (\text{resp.}\ \bar{I},\ \bar{I}s_4,\ \bar{I}s_4 s_5, \bar{I} s_5).
    \label{eq: }
\end{align}
The above operations are $\propto \bar{X}$ (resp.~$\propto \bar{I}$) by the GS [cf.~Eq.~\eqref{eq: gauge symmetry (logical X)}].

\paragraph{Horizontal}
The scheme for $\cz$ is obtained from
\begin{align}
    \cz_{\ctrl, \tgt} \propto & \prod^2_{i=0} \sum^1_{p=0} \left(\frac{I_\co I_\x I_\y + (-1)^p X_\co X_\x X_\y}{2}\right)_\paprime{4} \left(Z_\co Z_\x Z_\y\right)^p_\pa{2i}
    \label{eq: gauge symmetry of CZ (horizontal)}                                                                                                                                                                  \\
    =                   & \left(H_\x H_\y\cdot Z^{[01,10]}_\co\right)_\paprime{4}
    \\
                        & \quad \left(\prod^2_{i=0}\left(X^{[01,10]}_\co\right)_\pa{2i}\left(\sum^1_{p=0}(X_\co)^p_\paprime{4} (|p\rangle\langle p|_\co)_\pa{2i}\right)\left(X^{[01,10]}_\co\right)_\pa{2i}\right)
    \left(Z^{[01,10]}_\co\cdot H_\x H_\y\right)_\paprime{4} ,
\end{align}
where we have used Eq.~\eqref{eq: gauge symmetry (logical X)} for the control system and some key relations presented around Eqs.~\eqref{eq: derivation CX (first step)}--\eqref{q: derivation CX (key relation: third step (2))}.
The last line denotes sequentially applying the conditional operation $\sum^1_{p=0}(X_\co)^p_\paprime{4}(|p\rangle\langle p|_\co)_\pa{2i}$ for all particles $\{\pa{2i}|0\leq i\leq 2\}$.

With the ancillary particles initialized to $(|000\rangle_{\co,\po})\pa{j}$ for $j\in\{1,3\}$,
each of the above conditional operations consists of nearest-neighbor and next-nearest-neighbor CNOT gates as follows:
\begin{align}
    \sum_{p=0}^1(X_\co)_\paprime{4}(|p\rangle\langle p|_\co)_\pa{4} & : (|c_4\rangle_\co)_\pa{4} \xrightarrow{\mathrm{(i)}} (|c_4^\prime \oplus c_4\rangle_\co)_\paprime{4}
    \label{eq: CNOT 4 -> 4prime}
    \\
    \sum_{p=0}^1(X_\co)_\paprime{4}(|p\rangle\langle p|_\co)_\pa{2} & : (|c_2\rangle_\co)_\pa{2} \xrightarrow{\mathrm{(ii)}} (|c_2\rangle_\co)_\pa{3} \xrightarrow{\mathrm{(iii)}} (|c_4^\prime \oplus c_2\rangle_\co)_\paprime{4}
    \label{eq: CNOT 2 -> 3 -> 4}
    \\
    \sum_{p=0}^1(X_\co)_\paprime{4}(|p\rangle\langle p|_\co)_\pa{0} & :(|c_0\rangle_\co)_\pa{0} \xrightarrow{\mathrm{(ii)}}
    (|c_0\rangle_\co)_\pa{1} \xrightarrow{\mathrm{(iv)}} (|c_0\rangle_\co)_\pa{3} \xrightarrow{\mathrm{(iii)}} (|c_4^\prime \oplus c_0\rangle_\co)_\paprime{4}
    \label{eq: CNOT 0 -> 1 -> 3 -> 4}
\end{align}
where $c_i\in\{0,1\}$ ($i\in\{0,2,4\}$) and $c_4^\prime\in\{0,1\}$ denote spin states of the physical particles $\pa{i}$ and $\paprime{4}$,
and the procedures for the operations (i)--(iv) are given in the next paragraph.
Note that to initialize the spin states of ancillary particles after the conditional operation acting on $\paprime{4}$ and $\pa{2i}$ ($i\in\{0,1,2\}$),
we apply $(H_\co)_\pa{j}$ for all $j\in\{1,3\}$ to these particles and measure their spin state.
When $|1\rangle_\co$ is measured from either of them, the error $(Z_\co)_\pa{2i}$ occurs on the target system and we then update the Pauli frame of this system.

The operations labeled (i)--(iv) in Eqs.~\eqref{eq: CNOT 4 -> 4prime}--\eqref{eq: CNOT 0 -> 1 -> 3 -> 4} are realized as follows.
Here, each ancillary particle returns to the vertex labeled $00$ after each procedure.
First, the type-(i) procedure in Eq.~\eqref{eq: CNOT 4 -> 4prime} is the same as that discussed in the paragraph around Eqs.~\eqref{eq: fig/CX_1.pdf}--\eqref{eq: fig/CX_2.pdf}.
Second, the type-(ii) procedure in Eqs.~\eqref{eq: CNOT 2 -> 3 -> 4}--\eqref{eq: CNOT 0 -> 1 -> 3 -> 4} is realized through shuttling the ancillary particle $\pa{3}$ (resp.~$\pa{1}$)
as $X_\x \rightarrow X_\y\rightarrow X_\x \rightarrow X_\y$ with $(\cnot_\co)_{\pa{2},\pa{3}}$ (resp.~$(\cnot_\co)_{\pa{0},\pa{1}}$) applied before each shuttling operation.
Third, the type-(iii) procedure in the same equations is realized by a combination of the shuttling and tunneling operations for $\pa{3}$ and $\paprime{4}$, respectively, as
$(X_\y)_{\pa{3}}\rightarrow (X_\x)_\paprime{4} \rightarrow (X_\y)_{\pa{3}}$, with $(\cnot_\co)_{\pa{3},\paprime{4}}$ applied before each operator, illustrated as follows:
\begin{align}
    \begin{array}{c}
        \includegraphics[height=2.7cm]{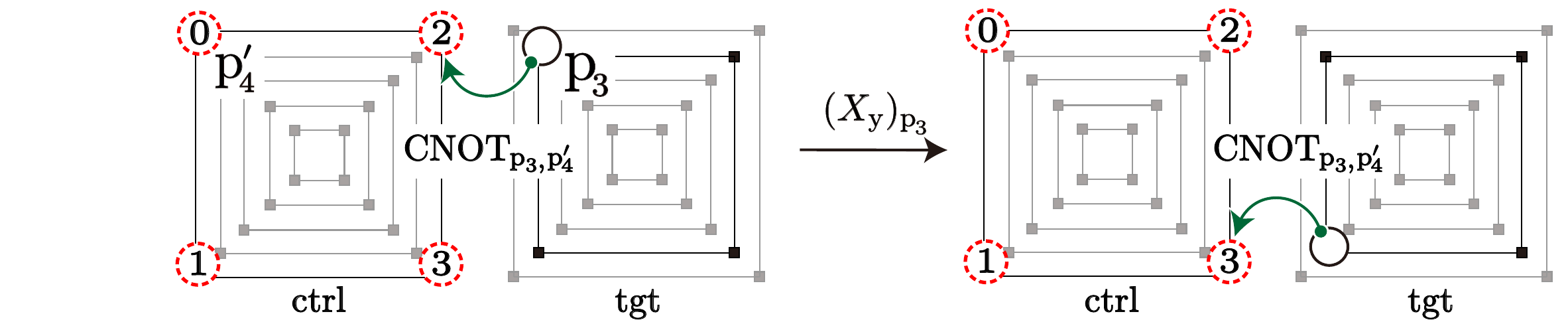},
    \end{array}
    \label{eq: fig/CZ_1.pdf}
    \\
    \begin{array}{c}
        \includegraphics[height=2.7cm]{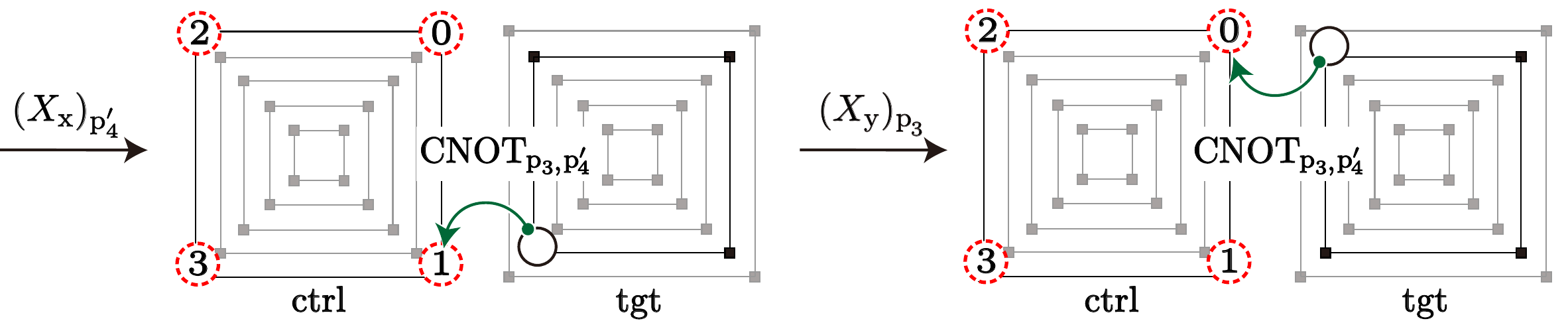}.
    \end{array}
    \label{eq: fig/CZ_2.pdf}
\end{align}
Finally, to realize the type-(iv) procedure, we simply apply the next-nearest-neighbor $(\cnot_\co)_{\pa{1},\pa{3}}$ to the vertices labeled 00.

\subsection{Remark on the GS in logical operation}\label{Subsection: }
Note that we have used only a small part of the GS for the logical operations,
leaving room for more significant architectural flexibility to emerge from a larger part of the GS.
Namely, we have demonstrated the stacking flexibility by exploiting only the GS under the stabilizer group
within the whole gauge algebra $\mathcal{G}$, as in Eq.~\eqref{eq: gauge symmetry (logical X)}.

The author suspects that such significant and interesting architectural flexibility will emerge when extending the QSD-based architectures toward fault-tolerant ones, 
as mentioned in the introduction, rather than in the present simple nested-square QEC model.
We stress that the present paper (and the companion paper~\cite{asaka2025quantum}) employs the simple QEC model
to explore the fundamental advantages offered by the integrated use of the QSD and the GS (and the quantum walk).

\section{Conclusion}\label{Section: Conclusion}
In this paper, we have demonstrated that the GS offers resilience against three types of noise inherent in QSD-based architectures.
We have also shown that the integrated use of the QSD and the GS offers a form of architectural flexibility,
i.e., stacking flexibility, which enables both vertical and horizontal scaling of the error-correcting system,
while the resource efficiency, the inherent advantage of our simple QSD-based nested-square system, is preserved.
The former is ensured by Eq.~\eqref{eq: gaugesymmetry (derivation of the generalized quantum noise)}.
In addition, the latter is ensured by Eq.~\eqref{eq: gauge symmetry (logical X)} together with the fact that
the implementation of both the stabilizer measurement and the three primitive logical gates ($\cx$, $\sqrt{\cx}$ and $\cz$) without QSD requires interactions beyond next-nearest-neighbor particles [cf.~Fig.~\ref{figure: fig/stabilizer_measurement}].

The present exploration, together with the companion paper~\cite{asaka2025quantum}, paves the way toward ultrafast fault-tolerant QSD-based computational architectures.
The resilience against the dominant noise inherent in the QSD-based architecture, offered by the GS, would be an essential ingredient for achieving fault-tolerance,
and the architectural flexibility offered by the integrated use of the QSD and the GS would be an essential aspect
to preserve the advantages of the QSD, e.g., parallel conditional gate operations, resource-efficiency, and time-independence (autonomy), as discussed in the introduction.
Here, quantum computation employing particles' spin and position states constitutes a general quantum many-body system,
and thus its study would yield crucial insights for quantum many-body physics. 
Indeed, the present study indicates that the stabilizer formalism is a powerful tool for describing the dynamics of multiple particles under nontrivial noise, 
each having both spin and position degrees of freedom.

\section*{Acknowledgment}
The author thanks anonymous referees of the previously submitted Ref.~\cite{asaka2025quantum}, as their comments inspired the present work.
This work was supported by JSPS KAKENHI Grant Number 26K17055.

\appendix
\section{Simple Estimate of Failure Probability}\label{Section: Simple Estimate of Failure Probability}
As shown in Sec.~\ref{Subsection: Stabilizer measurement} and Fig.~\ref{figure: fig/stabilizer_measurement},
exploiting the QSD, or more precisely, encoding a three-qubit state into the spin and position states of a single particle,
increases the circuit depth of the stabilizer measurement compared to a common implementation where a single particle carries only a single-qubit state, as a tradeoff for the stacking flexibility.

Nevertheless, we here argue that the increase in the failure probability of the stabilizer measurement may be marginal, as shown by a simple estimation.
Here, the analysis is based on the following assumptions:
\begin{enumerate}
    \item In both implementations, we define the failure as the event that some errors affect at least two particles after one cycle of stabilizer measurement.
          For the QSD-based implementation, some errors affecting at least two states in a single particle also cause the failure.
    \item Basically, any gate operation for the stabilizer measurement causes an error with probability $p$.
          For the QSD-based implementation, a gate operation acting on one of the states causes an error that acts only on this state.
    \item The following gate operations appearing in Eqs.~\eqref{eq: stabilizer measurement (impl 1)}--\eqref{eq: stabilizer measurement (impl 3)}
          are treated as a single gate operation since the physical particle $\pa{j\pm 1}$ effectively occupies only one vertex of its square:
          \begin{align}
              \left((X_\y)_{\pa{j}} (\cnot_\co)_{\pa{j\pm 1},\pa{j}} (X_\x)_{\pa{j}} (\cnot_\co)_{\pa{j\pm 1},\pa{j}}\right)^2,
              \\
              \left((X_\y)_{\pa{j}} (\cnot_\co)_{\pa{j},\pa{j\pm 1}} (X_\x)_{\pa{j}} (\cnot_\co)_{\pa{j},\pa{j\pm 1}}\right)^2.
              \label{eq: }
          \end{align}
          Indeed, in the above operations, a nearest-neighbor CNOT is applied effectively only once to physical particle $\pa{j\pm 1}$ ($j\in\{1,3\}$) at a vertex.
\end{enumerate}
As future work, we should refine the above estimation mainly in two aspects: (i) errors of an ancillary particle $\pa{j}$ are not accounted for,
and (ii)  it does not differentiate between the flipping operation ($X_\co^{[\{xy\}]}$ or $Z_\co^{[\{xy\}]} $) and the CNOT operation that would be more costly than the former.
Both refinements would depend on the implementation platform.

First, the failure rate of the common implementation, which assigns one particle (qubit) to each horizontal line on the circuit in Fig.~\ref{figure: fig/stabilizer_measurement}~(a),
is given by
\begin{align}
    E_\textrm{common}   & = 1-  p_{\textrm{none}} - p_{\textrm{single}} = 341p^2 + O(p^3)
    \\
    p_{\textrm{none}}   & = \left(1-p^{(2)}\right)^4 \left(1-p^{(3)}\right)^2 \left(1-p^{(4)}\right)^2 \left(1-p^{(6)}\right)
    \\
    p_{\textrm{single}} & = \left(\frac{4p^{(2)}}{1-p^{(2)}}+\frac{2p^{(3)}}{1-p^{(3)}} + \frac{2p^{(4)}}{1-p^{(4)}} + \frac{p^{(6)}}{1-p^{(6)}}\right)p_{\textrm{none}},
    \label{eq: }
\end{align}
where $p^{(n)} := 1-(1-p)^n$ is the error rate of a particle undergoing $n$ gate operations.
The above estimation follows from the fact that four, two, two, and one particles undergo two, three, four, and six gate operations (CNOT gates), respectively.

Second, the failure rate of the present QSD-based implementation is given by
\begin{align}
    E_\textrm{present}         & = 1 - p^\prime_{\textrm{none}} - p^\prime_{\textrm{single}} =537p^2 + O(p^3),                                                                                   \\
    p^\prime_{\textrm{none}}   & = (1-\bar{p}_\co)(1-p_\co)^2(1-p_\x)^3(1-p_\y)^3,                                                                                                 \\
    p^\prime_{\textrm{single}} & =      \left(\frac{\bar{p}_\co}{1-\bar{p}_\co}+\frac{2p_\co}{1-p_\co}+\frac{3p_\x}{1-p_\x} + \frac{3p_\y}{1-p_\y}\right)p^\prime_{\textrm{none}}.
\end{align}
Here, $p_\co := p^{(7)}$ (resp.~$\bar{p}_\co:=p^{(10)})$ is the error rate of the spin state of physical particle $\pa{0}$ or $\pa{4}$ (resp.~$\pa{2}$)
that undergoes four spin-flip or phase-flip operations and three (resp.~six) CNOT operations.
In addition, $p_\x$ and $p_\y$ are the error rates of the $x$-axis and $y$-axis position states of any physical particle that undergoes Hadamard gates $H_\x$ and $H_\y$ twice.

\begin{figure}[t]
    \centering
    \begin{minipage}[b]{0.99\columnwidth}
        \centering
        \includegraphics[width=14.8cm]{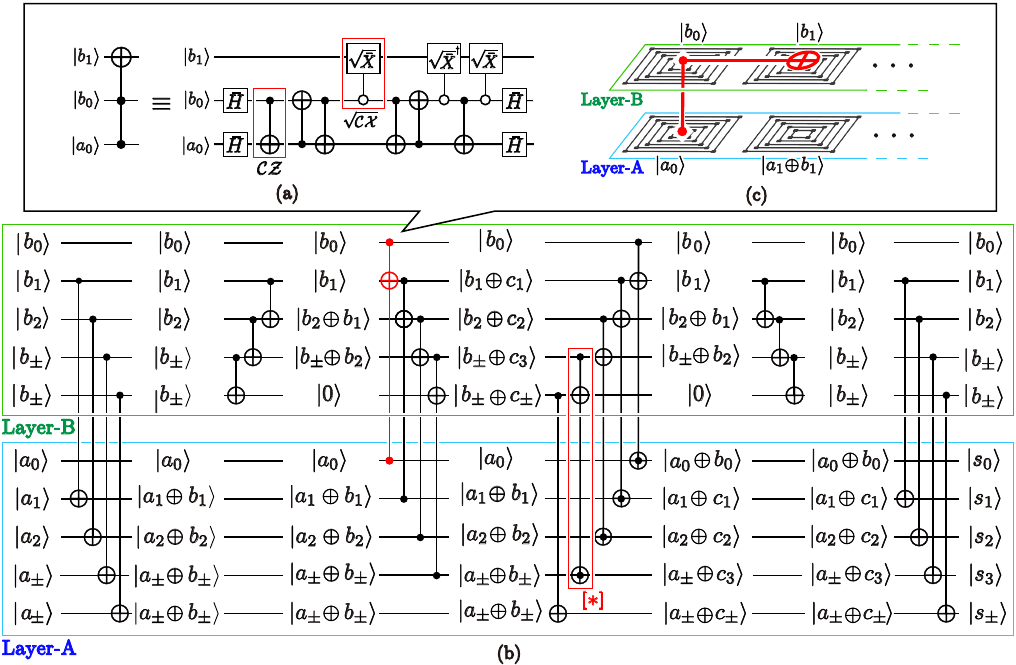}
        \caption{
        (a) Circuit for the Toffoli gate, which applies logical $\bar{X}$ operator to the top system if and only if both of the other two systems are in the $-1$ eigenstate of $\bar{Z}$.
        (b) Quantum adder~\cite{thapliyal2013design, thapliyal2016mapping, asaka2020quantum} that adds two numbers $a$ $(=a_2a_1a_0)$ and $b$ $(=b_2b_1b_0)$ and outputs their sum $s=a+b$ $(=s_3s_2s_1s_0)$, where $a_i,b_i,s_i\in\{0,1\}$ for $0\leq i\leq 3$.
        The signs of their numbers are denoted by $a_\pm$, $b_\pm$, and $s_\pm$, each taking the value $0$ or $1$ to indicate positive or negative sign, respectively.
        Here, $c_j$ is defined by $c_1 := a_0 b_0$, $c_j:=a_{j-1}b_{j-1} \oplus b_{j-1} c_{j-1} \oplus c_{j-1} a_{j-1}$ for $2\leq j\leq 3$, and $c_\pm:=a_\pm b_\pm \oplus a_\pm c_3 \oplus b_\pm c_3$.
        Similarly, we define $s_0 := a_0\oplus b_0$, $s_i:=a_i\oplus b_i\oplus c_i$ for $1\leq i\leq 2$, and $s_3:=a_\pm \oplus b_\pm \oplus c_3$.
        The gate labeled [*] in the circuit consists of the Toffoli gate followed by CNOT gate.
        For a more detailed discussion of this circuit, see \cite{asaka2020quantum}.
        (c) Two-layer architecture for linearly arranged three-particle systems.
        Our QEC system enables a space-efficient implementation of the quantum adder in such an architecture.
        The key point is that $\cx$, $\sqrt{\cx}$, and $\cz$ can be implemented between vertically or horizontally aligned systems,
        with the required interactions extending no farther than next-nearest neighbors.
        }
        \label{figure: adder}
    \end{minipage}
\end{figure}

\bibliography{ref}

\end{document}